\documentclass[footinbib,PRB,twocolumn,amsmath,amssymb,floatfix]{revtex4-1}
\usepackage{epsfig,psfrag,subfigure}
\usepackage{colordvi}

\def\bm#1{\mbox{\boldmath$#1$}}

\def\op{\eta}
\def\opmag{\vert \eta \vert}
\def\posvec{{\bf r}}
\def\coeffgrad{\mathcal{K}}  
\def\coeffquad{\mathcal{A}}  
\def\coeffquart{\mathcal{B}}   
\def\matI{I} 
\def\matE{E} 
\def\Ldw{L}
\def\coeffmirror{\tau} 
\def\vort{W} 
\def\jun{{\cal J}}
\def\sfv{V}

\def\dwrhodown{\rho_\text{dw}}
\def\vrho{\rho_\text{v}}

\def\scalegrad{\tilde{\coeffgrad}}
\def\scalequart{\tilde{\coeffquart}}
\def\mirror{M}

\def\Hdual{\Lambda}

\def\fdwcore{f_\text{core}}

\def\bendangle{\Theta}

\def\degensubspace{\mathcal{R}}
\def\Phibend{\Phi_\text{bend}}
\def\Phivarbend{\Phi_\textrm{var,bend}}
\def\dgfGamma{\Gamma}

\begin{document}

\title{Penetration of nonquantized magnetic flux through a domain-wall bend in time-reversal symmetry broken superconductors}

\author{David G.~Ferguson and Paul M.~Goldbart }

\affiliation{Department of Physics,
Institute for Condensed Matter Theory,
and Frederick Seitz Materials Research Laboratory,\\
University of Illinois at Urbana-Champaign, Urbana, IL 61801, USA}
\date{\today}

\begin{abstract}
It has been proposed that the superconductivity of Sr$_2$RuO$_4$ is characterized by pairing that is unconventional and, furthermore, spontaneously breaks time-reversal symmetry.  However, one of the key expected consequences, viz., that the ground state should exhibit chiral charge currents localized near the boundaries of the sample, has not been observed, to date.  We explore an alternative implication of time-reversal symmetry breaking: the existence of walls between domains of opposing chirality.  Via a general phenomenological approach, we derive an effective description of the superconductivity in terms of the relevant topological variables (i.e., domain walls and vortices). Hence, by specializing to the in the in-plane rotationally invariant limit, we show that a domain wall that is translationally invariant along the $z$ axis and includes a bend through an angle $\Theta$ is accompanied by a nonintegral (and possibly nonquantized) magnetic {\it bend flux} of $\big((\Theta/\pi)+n\big)\Phi_0$, with integral $n$, that penetrates the superconductor, localized near the bend.  We generalize this result to the situation in which gauge transformations and rotations about the $z$ axis are degenerate transformations of the chiral superconducting order.  On the basis of the specialized result and its generalization, we note that any observation of localized, nonquantized flux penetrating a $z$-axis surface (e.g., via scanned-probe magnetic imaging) can be interpreted in terms of the presence of bent walls between domains of opposing chirality, and hence is suggestive of the existence of time-reversal symmetry-breaking superconductivity.
\end{abstract}
\maketitle

\section{Introduction}
Recently, there has been developing excitement regarding the nature of the superconducting state of the crystalline compound Sr$_2$RuO$_4$~\cite{Mackenzie03}.  This is because, as in superfluid $^3$He, the superconductivity has been proposed to be unconventional, having Cooper pairs of the triplet type~\cite{Rice95,Ishida98,Nelson04}.  In addition, recently obtained evidence for the existence of half-quantum vortex structures~\cite{Jang10}, which are expected to support zero-energy Majorana modes~\cite{Kopnin91,Read00}, suggests that Sr$_2$RuO$_4$ could potentially be used as a host medium for topological quantum computing~\cite{Kitaev03,Nayak08}.  However, questions remain concerning the structure of the pairing state (see, e.g., Refs.~\cite{Yoshioka09,Raghu10}) and, in particular, whether the superconductivity does indeed spontaneously break time-reversal symmetry, and would thus form a chiral state~\cite{Luke98,Xia06,Kallin09}.  In particular, the theoretical prediction (see e.g.~\cite{Matsumoto99}) that the ground state should exhibit chiral charge-currents localized near the boundaries of the sample, has not been verified experimentally, to date, despite considerable efforts~\cite{Bjornsson05,Kirtley07,Hicks10}.  Moreover, if time-reversal symmetry were broken by the superconductivity of Sr$_2$RuO$_4$ then---in addition to vortices---domain walls that separate regions of opposing chirality would enter as a new topological feature of the theory~\cite{Volovik85}.

For {\it conventional} superconductivity, the phenomenological approaches of London~\cite{London35} and of Ginzburg and Landau~\cite{Ginzburg50} predated the formulation of the microscopic theory, due to Bardeen, Cooper, and Schrieffer~\cite{Bardeen57}.  For {\it unconventional} superconductors, including those in which time-reversal symmetry is spontaneously broken, it is likewise possible to make progress phenomenologically, without invoking detailed information about any specific microscopics~\cite{Volovik85,Sigrist91,Mineev99}.  This is the approach that we adopt in the present paper, as we explore certain specific features of time-reversal symmetry-broken states: (i)~the possibility that there are walls between domains of opposing chirality, (ii)~the threading of these walls by magnetic flux, and (iii)~the fact that this flux may penetrate in nonquantized amounts~\cite{Sigrist89,Sigrist99}.  Lack of flux quantization has been discussed in related settings, such as superfluid condensates of ionized hydrogen~\cite{Babaev04}, as well as time-reversal symmetry-broken superconductors that feature spin-polarization~\cite{Volovik84}, disclinations~\cite{Volovik00}, or intersecting grain-boundaries between crystallites~\cite{Sigrist95}.

Our central result is as follows: nonintegral (and even nonquantized) multiples of the superconducting quantum of magnetic flux penetrate time-reversal symmetry-breaking superconductors, localized near bends in walls between chiral domains.  We first obtain this result via an effective description in terms of domain walls and vortices, which shows that (in the special case of the crystallographically in-plane rotationally invariant limit) a domain wall that is translationally invariant along the $z$ axis and bends through an angle $\Theta$ is accompanied by a net flux (which we term ``bend flux'') of $\big((\Theta/\pi)+n\big)\Phi_0$, localized in the bend region, for arbitrary integral $n$.  We then generalize this result to the situation in which gauge transformations and rotations about the $z$ axis are degenerate transformations of the maximally chiral superconducting order (i.e., are transformations that have equivalent impacts).  If the rotational symmetry is broken down to discrete tetragonal symmetry, our central result remains valid for the particular case of a domain wall bent through $\pi/2$ radians.

Our specialized and more general results indicate that observations of localized, nonquantized flux penetrating a $z$-axis surface (e.g., via scanned-probe magnetic imaging), could potentially be interpreted in terms of the presence of bent walls separating domains of opposing chirality, and hence would be suggestive of the existence of time-reversal symmetry-breaking superconductivity. Alternatively, if localized nonquantized flux is not observed to penetrate a $z$-axis surface, this would suggest that either (i) domain walls are not present, or (ii) domain walls are present, but are arranged in a parallel array and thus are not bent.

This paper is organized as follows.  In Section~\ref{sec:PhenomUSC} we review the structure of the Ginzburg-Landau order parameter appropriate for unconventional superconductivity with broken time-reversal symmetry, along with the corresponding Ginzburg-Landau free-energy functional.  In Section~\ref{sec:Top_field_configurations} we analyze this free energy via an extension of the London limit, in which we exchange the Ginzburg-Landau order-parameter description for a reduced description in terms of the collection of spatially varying  \lq\lq phase-like\rq\rq\ fields that for homogeneous configurations would parametrize the space of equilibrium states.  The extension amounts to taking the limit in which the domain walls are vanishingly thin, compared with the London penetration depth. (It should be noted that Heeb and Agterberg, in Ref.~\cite{Heeb99}, attach a different meaning to the term \lq\lq extended London limit\rlap{\rq\rq}.\thinspace)\thinspace\  Taking this limit enables us to focus on the structure and implications of the topological excitations of the order parameter, which have the form of vortices and domain walls, and to relate the densities of these excitations to singularities in the phase-like fields.  In Section~\ref{sec:effective_theory} we return to the free energy, expressing it in terms of these excitation densities, and in Section~\ref{sec:bending_domain_wall} we use this framework to determine the spatial distribution of magnetization associated with domain-wall topological excitations, specifically walls that contain bends.  In this section we also show that the thin-domain-wall limit is not essential, in the sense that the key features of our results, such as the threading of bent domain walls by nonquantized amounts of magnetic flux, continue to hold, even when this limit is relaxed.  In Section~\ref{sec:Experiments} we consider three experimental settings in which nonquantized flux may be observed; positive results in any one of them would provide evidence for the existence of time-reversal symmetry-breaking superconductivity.  Finally, in Section~\ref{sec:conclusion} we summarize our key results and their implications.  Some technical details are relegated to a pair of appendices.

\section{Phenomenological theory of unconventional superconductivity}
\label{sec:PhenomUSC}

In this section we describe the phenomenological theory of superconductivity, on which our analysis is rooted.
This approach is based on the notion of an appropriate superconducting order parameter, along with general symmetry considerations, and thus can be explored independently of any specific microscopic details.  The order parameter transforms under the full symmetry group of the physical system, and thus provides a representation of this symmetry group.  (It is common, in the context of planar superconductors, for the point-group aspect of this symmetry to be the tetragonal group $D_{4h}$, reflecting the underlying electronic and atomic structure of the crystalline material.)\thinspace\ For $\textrm{Sr}_2 \textrm{RuO}_4$, the material on which we shall focus, it is known that, for a range of temperatures close to the superconducting transition temperature $T_{\rm c}$, the superconducting properties are nearly isotropic with respect to rotations about the $z$ axis~\cite{Mao00,Deguchi04} (i.e., the direction perpendicular to the RuO$_2$ planes), and is only weakly tetragonal about this axis.  Although, accordingly, the initial focus of our analysis will be on the isotropic limit (which we term the in-plane rotationally or ${\rm SO(2)}_{z}$-invariant limit), we do subsequently address the cases in which the symmetry is lowered to the discrete group $D_{4h}$ (and also, parenthetically, the group $D_{6h}$).  At the outset, we therefore retain generality by determining the representation furnished by the superconducting order parameter appropriate to $D_{4h}$ symmetry, motivated by the relevance of this group to $\textrm{Sr}_2 \textrm{RuO}_4$.

We now determine the appropriate representation of the superconducting order parameter, bearing in mind the foregoing symmetry considerations.  This choice of representation is made according to the following three simplifying assumptions: (i)~The ground state of the superconducting order should transform trivially under lattice translations.  Thus, at the lengthscales relevant for a phenomenological description such as the one used here, the ground state of the superconducting order is translationally invariant.  (ii)~The representation of the symmetry group should be irreducible.  This is justified in the case of $\textrm{Sr}_2\textrm{RuO}_4$ as, in the absence of an applied magnetic field, only one superconducting transition seems to be observed.  (Recent experiments on Sr$_2$RuO$_4$ under uniaxial pressure do, however, indicate the possibility of a second transition \cite{Kittaka10}.)\thinspace\ (iii)~The representation should allow for the possibility that the superconducting state spontaneously breaks time-reversal symmetry.  This would require that the dimension of the representation be greater than unity.  Taken together, these assumptions fix the order parameter to transform according to the $\Gamma_5$ representation~\cite{Sigrist91}, which is two-dimensional~\footnote{We do not need to specify whether the representation is $\Gamma_5^+$ or $\Gamma_5^-$, for which the basis functions transform respectively as $\{XZ,YZ\}$ or $\{X,Y\}$.  The results of the present work apply to both cases.}.  Accordingly, the order parameter is the complex-valued, two-component field $\op_{a}(\posvec)$, where the index $a$ runs through the corresponding basis functions of the representation (i.e., $X$ and $Y$) which, in general, depends on the three-dimensional position vector $\posvec$.  To simplify our analysis, we consider superconducting states that are translationally invariant along the $z$ axis, thus rendering the physical problem effectively two-dimensional.  Provided we apply external magnetic fields that are oriented along the $z$ direction (i.e., ${\bf H}=H\hat{\bf z}$), this is an option, owing to the intrinsic translational invariance of the system along the $z$ direction.  These requirements, taken together, then dictate that the appropriate Ginzburg-Landau free energy functional governing the $\Gamma_5$ representation is given by~\cite{Volovik85,Sigrist91}
\begin{equation}
\begin{split}
    \label{eqn:F_GLGamma5}
    &F'[\op']
    =
    \int d^2r'
    \big\{
    \coeffgrad'_{a b c d} \,
    (D'_a \op'_b)^\ast
    (D'_c \op'_d)
    -\coeffquad \, {\op}_a^{\prime \ast}\,{\op}_a^{\prime \phantom{\ast}}
    \\
    &\quad+\frac{1}{2}\coeffquart'_{a b c d} \,
    {\op}_a^{\prime \ast} \,
    {\op}_b^{\prime \ast} \,
    {\op}_c^{\prime \phantom{\ast}}\,
    {\op}_d^{\prime \phantom{\ast}}
    + \frac{1}{8\pi}\vert({\bm \nabla}'\times{\bf A}')- {\bf H}'\vert^2
    \big\}.
\end{split}
\end{equation}
Here, two-dimensional summations are implied over the repeated indices $a,b,c,d$, and the covariant derivative is defined via ${\bm D}' := {\bm \nabla'} - 2 \pi i {\bm A'}/\Phi_0$, where $\Phi_0$ is the superconducting flux quantum $hc/2e$.

In Eq.~(\ref{eqn:F_GLGamma5}) the primed variables are dimensionful. We now define relevant scales of length and energy, and use them to introduce convenient dimensionless variables, which we use throughout the remainder of the paper and which we write without primes.  As a first step, for the coefficient tensors $\coeffgrad'$ and $\coeffquart'$ we define the dimensionful scale factors $\scalegrad$ and $\scalequart$, which we then use to construct the dimensionless tensors $\coeffgrad:= \coeffgrad' / \scalegrad$ and $\coeffquart:= \coeffquart' / \scalequart$.  In the in-plane rotationally invariant limit, symmetry considerations dictate that $\coeffgrad$ and $\coeffquart$ can be parametrized in the following way:
\begin{equation}
\label{eqn:TensorParameters}
\begin{split}
    \coeffquart_{abcd}
    &=
    \matI_{ac}\matI_{bd}
    +
    \frac{\sigma}{2} M^\delta_{ac} M^\delta_{bd},
    \\
    \coeffgrad_{abcd}
    &=
    \matI_{ac}\matI_{bd}
    +
    \mu \matE_{ac} \matE_{bd}
    +
    \frac{\tau}{2} M^\delta_{ac} M^\delta_{bd},
\end{split}
\end{equation}
where a summation from 1 to 2 is implied over the repeated index $\delta$, the three real parameters $\{\sigma,\mu,\tau\}$  are, in principle, temperature dependent, and the constant tensors
$\{{\bm \matI}, {\bm \matE},{\bm \mirror}^{1},{\bm \mirror}^{2}\}$
are defined via
\begin{eqnarray}
\begin{split}
 {\bm \matI}&:=\left(\begin{array}{cc} 1 & 0 \\ 0 & 1 \\ \end{array} \right),
 \qquad\quad\,\,\,\,
 {\bm \matE}:=\left(\begin{array}{cc} \phantom{-}0 & 1 \\ -1 & 0 \\ \end{array} \right),
 \\
 {\bm \mirror}^{1}&:=\left(\begin{array}{cc} 1 & \phantom{-}0 \\ 0 & -1 \\ \end{array} \right),
 \qquad
 {\bm \mirror}^{2}:=\left(\begin{array}{cc} 0 & 1 \\ 1 & 0 \\ \end{array} \right).
\end{split}
\end{eqnarray}
The choice of tensor decomposition in Eq.~(\ref{eqn:TensorParameters}) is motivated by the observation that under $\text{SO}(2)_z$ rotations ${\bm \matI}$ and ${\bm \matE}$ each transform trivially, whereas the pair
$\{{\bm \mirror}^{1},{\bm \mirror}^{2}\}$ mix.  If the symmetry were not $\text{SO}(2)_z$ but $\text{D}_{4h}$, the coefficients of the the ${\bm \mirror}^{1}{\bm \mirror}^{1}$ and ${\bm \mirror}^{2}{\bm \mirror}^{2}$ terms may be unequal; however, for $D_{6h}$ symmetry they would remain the same.

We choose a natural scale for the order parameter, viz.,
$\op_0 = (\coeffquad/\scalequart)^{1/2}$,
and use it make the definition $\op:= \op'/\op_0$.
We then define the two lengthscales:
(i)~the penetration depth $\lambda:= \Phi_0 /(32 \pi^3 \op_0^2 \scalegrad )^{1/2}$,
which characterizes the lengthscale for variations of the magnetic field;
and (ii)~the coherence length $\xi:= (\scalegrad / \coeffquad)^{1/2}$,
which characterizes the lengthscale for variations in the amplitude of the order parameter.
We then scale all lengths by $\lambda$, via
$(\posvec,{\bm \nabla},{\bm D}):=(\posvec'/\lambda,\lambda {\bm \nabla}',\lambda {\bm D}')$.
We also make the conventional definition of the Ginzburg-Landau parameter
$\kappa:= \lambda/\xi$.
Next, we define the dimensionless
vector potential ${\bm A}$,
applied magnetic field ${\bm H}$, and
magnetic flux $\Phi$ via
$({\bm A},{\bm H},\Phi):=(2 \pi \lambda {\bm A}'/\Phi_0,2 \pi \lambda^2{\bm H}'/\Phi_0,2 \pi \Phi' /\Phi_0)$.
We note that, with this choice of units, a flux equal to a flux quantum has the dimensionless value $2\pi$.
As a final step in the construction of the dimensionless variables we choose as a scale for the free-energy density the value $f_0 = 2 \scalegrad \op_0^2/\lambda^2$, using which we arrive at the dimensionless free energy via $F:=F'/\lambda^2 f_0$.
It will be convenient for us to separate contributions to the dimensionless free-energy density into two groups:
the ``London type'' terms $f_{\rm L}$ and the and potential terms $f_{\rm P}$, respectively defined via
\begin{subequations}
\label{eqn:dimnesionlessGL}
\begin{eqnarray}
    f_{{\rm L}}\!&=&\!
    \frac{1}{2} \coeffgrad_{abcd}
    (D_a {\op}_b)^\ast
    (D_c {\op}_d)
    + \frac{1}{2}\vert({\bm\nabla}\times{\bf A})- {\bf H}\vert^2,
    \\
    f_{{\rm P}}\!&=&\!
    \frac{1}{2}\kappa^2
    (
    -{\op}_a^\ast {\op}_a^{\phantom{\ast}}
    +\frac{1}{2} \coeffquart_{abcd}
    \op_a^\ast \,
    \op_b^\ast \,
    \op_c^{\phantom{\ast}}\,
    \op_d^{\phantom{\ast}}
    ),
    \label{eqn:potential}
\end{eqnarray}
so that
\begin{equation}
F[\op] = \int d^2r \{ f_{{\rm L}} + f_{{\rm pot}} \}.
\end{equation}
\end{subequations}

One way in which $F$ differs from the conventional Ginzburg-Landau free energy is that the tensors $\coeffgrad$ and $\coeffquart$ allow for a larger number of material-dependent parameters, the latter free energy having only a single such parameter, viz., $\kappa$.  It is possible to estimate these additional parameters under the assumptions of weak coupling and a cylindrical Fermi surface~\cite{Zhu97}, and this results in the following values: $(\mu,\tau,\sigma)=(0,1,1)$.
However, due to the presence in ${\rm Sr}_2 {\rm RuO}_4$ of effects such as
multiple electronic energy bands~\cite{Bergemann00},
spin-orbit interactions~\cite{Haverkort08},
and chiral currents~\cite{Matsumoto99},
the parameters of a Ginzburg-Landau theory that incorporates such effects self-consistently are expected to be modified from their weak-coupling values, perhaps significantly.
Thus, we shall not limit our analysis to the weak-coupling values of these parameters.

In the section that follows, we analyze the potential terms of the phenomenological free energy, Eq.~(\ref{eqn:potential}), and, specifically, review how its structure leads to both vortices and domain walls.  In particular, we derive the vortex and domain-wall densities in terms of the \lq\lq phase-like\rq\rq\ variables; in the subsequent section, Section~\ref{sec:effective_theory}, we construct the effective free energy in terms of topological variables, such as the vortex and domain-wall densities.

\section{Topological field configurations}%
\label{sec:Top_field_configurations}%
As is well known, for many purposes, the state of an ordered phase can be adequately specified via an order-parameter field that takes values lying in the subspace of degenerate homogeneous equilibrium states~$\degensubspace$ (see, e.g., Ref.~\cite{Mermin79}).  If, as an example, different regions of a sample were to adopt distinct such values, it can---depending on the structure of the order parameter---be possible for the system to become trapped into order-parameter configurations that possess topologically stable defects~\cite{Mermin79}.  These are spatially varying configurations of the order parameter that cannot be removed via local deformations. The framework of homotopy groups of $\degensubspace$ enables one to identify and classify the possible topologically stable defects.

As is also well known (see, e.g., Refs.~\cite{Jackiw76,Su80,Read00,Ran09}), there can be a rich interplay between the topological features of the (bosonic) order-parameter fields that describe ordered phases and the qualitative character of any fermionic particles moving in the presence of such order-parameter fields.  However, in the present work we shall only consider the topological features of the appropriate order-parameter field, leaving for future work the analysis of its implications for the motion of fermions.

\begin{figure}
\includegraphics[width= .5\textwidth]{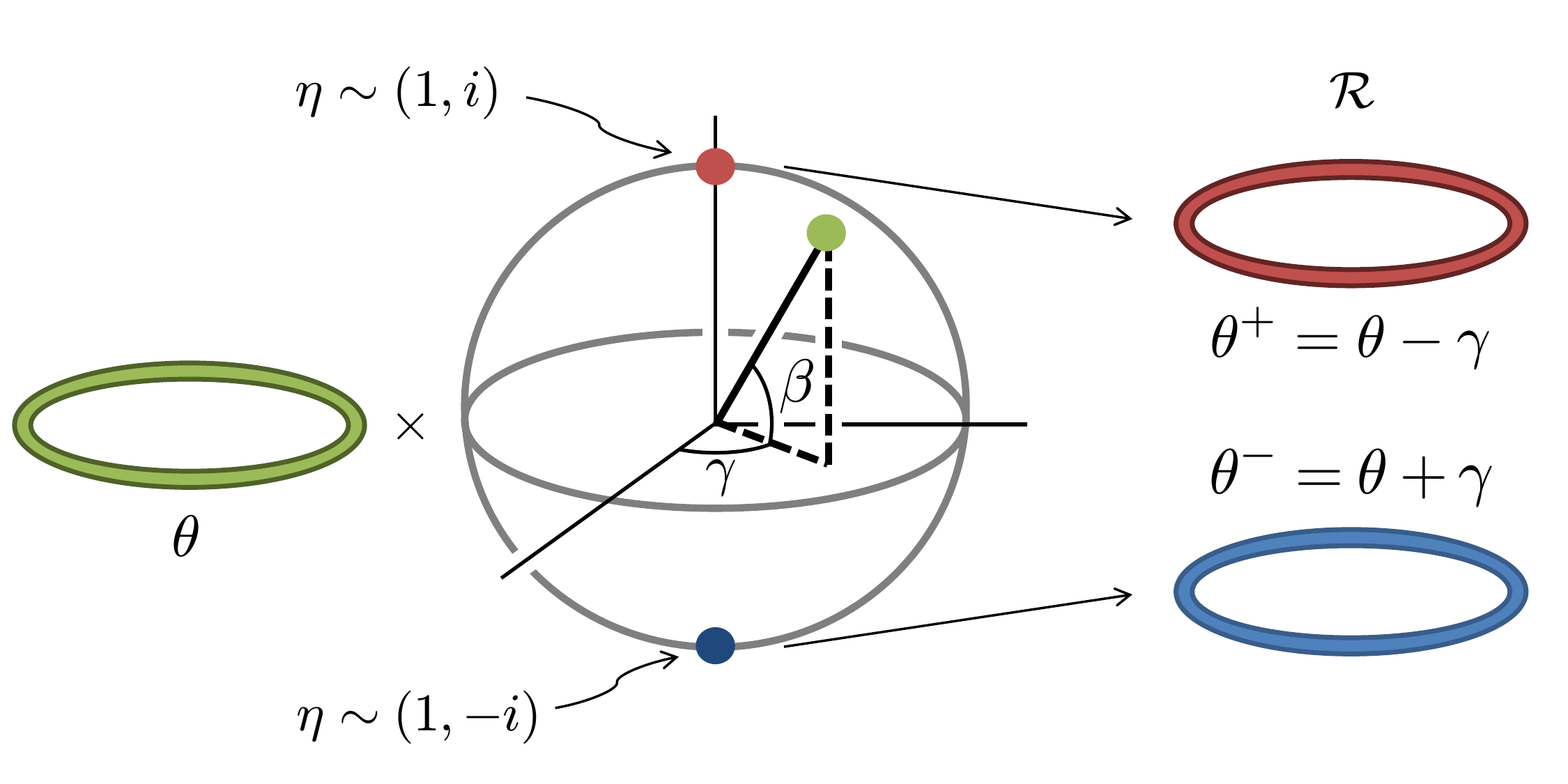}
\caption{
Visualizations of the order parameter space.  In the standard London limit, in which $\kappa \rightarrow \infty$, the order parameter is restricted to the coset space of configurations $S^1 \times S^2 /\mathbf{Z}_2$ [see the discussion following Eq.~(\ref{eqn:fpot})].  When, in addition, the Ginzburg-Landau parameter $\sigma$ of Eq.~(\ref{eqn:TensorParameters}) is positive, the north and south poles of the sphere become free-energy minima, and are thus energetically preferred, relative to the equator.  In the thin domain-wall limit (i.e., $\Ldw\to 0$), which we term the extended London limit, this preference is extreme.  In this latter case, the coset space describing degenerate minimum-energy configurations becomes $S^1 \times \{+,-\}$. (color online)}
\label{fig:orderparameter_visual}
\end{figure}

To determine $\degensubspace$ for the present problem, we follow the standard approach (see, e.g., Ref.~\cite{Sigrist91}) and analyze the structure of the potential terms of Eq.~(\ref{eqn:F_GLGamma5}). To simplify the analysis, it is useful to parametrize the two complex fields of the superconducting order parameter
${\bm \op} = (\op_X , \op_Y)$
in terms of four real fields
$\{\opmag,\theta,\gamma,\beta\}$
that transform simply under the operations of the symmetry group:
\begin{subequations}
\begin{eqnarray}
\label{eqn:op_def}
 {\bm \op}
    &&=
    \opmag e^{i \theta } {\bm R}^\gamma
    \cdot
    \left( \begin{array}{c}
    \phantom{i}\cos(\beta/2) \\ i \sin(\beta/2) \\ \end{array} \right),
    \\
    {\bm R}^\gamma
    &&:=
    \left(
    \begin{array}{cc}
    \cos \gamma & -\sin \gamma \\
    \sin \gamma & \phantom{-} \cos \gamma  \\
    \end{array}
    \right).
\end{eqnarray}
\end{subequations}
Now, elements of the product group ${\rm U}(1) \times {\rm SO}(2)_z$
of gauge transformations and $z$-axis rotations can be parametrized
via a phase angle $\theta'$ and a rotation angle $\gamma'$.
Under such elements, the order parameter ${\bm \op}$ transforms as
\begin{equation}
\label{eqn:param_transformations_notT}
    {\bm \op}(\opmag,\theta,\gamma,\beta)
    \rightarrow
    {\bm \op}(\opmag,\theta+\theta',\gamma+\gamma',\beta);
\end{equation}
under time reversal, ${\bm \op}$ transforms as
\begin{equation}
\label{eqn:param_transformations_T}
    {\bm \op}(\opmag,\theta,\gamma,\beta)
    \rightarrow
    {\bm \op}(\opmag,-\theta,\gamma,-\beta).
\end{equation}
Thus, we see that the parametrization of ${\bm \op}$, Eq.~\ref{eqn:op_def}, is given in terms of an amplitude $\opmag$ and phase $\theta$ that are similar to those used in conventional superconductivity, but also two angular variables, $\gamma$ and $\beta$, that respectively characterize the additional nontrivial ${\rm SO}(2)_z$ and time-reversal structure associated with the version of  unconventional order under consideration (but see~\footnote{This choice of parametrization is similar to that used in Ref.~\cite{Sigrist99}, in which the aforementioned additional structure of the order is parametrized by the scalar fields $\alpha$ and $\chi$ via ${\bm \eta} \propto {\bm R}^{-\alpha / 2} \cdot \big( \cos(\chi +\pi/4),i \sin(\chi+\pi/4)\big)$.}).
In terms of the parametrization given in Eq.~(\ref{eqn:op_def}), the potential terms~(\ref{eqn:potential}) become
\begin{equation}
\label{eqn:fpot}
    f_\text{P} = - \frac{\kappa^2}{2} \opmag^2 + \frac{\kappa^2}{4} \opmag^4 + \frac{1}{8 \Ldw^2} \opmag^4 \cos^2 \beta,
\end{equation}
in which we have introduced the dimensionless length $\Ldw = \sigma^{-1/2} / \kappa$, which will turn out to determine the domain-wall width.  As required by ${\rm SO}(2)_z$ and time-reversal invariance, these potential terms are independent of $\gamma$, as well as being even functions of $\beta$.  If the symmetry were reduced to $D_{4h}$, there would be the possibility of an additional term, proportional to $\vert \op \vert^4 \cos (4 \gamma) \cos^2 (\beta) $. In the present setting, to achieve the standard London limit, in which the magnitude of the order parameter $\opmag$ is fixed at unity, we take the {\it joint} limit $(\kappa,\sigma)\rightarrow(\infty,0)$, keeping $\Ldw$ fixed.  In this limit, the structure of the order-parameter space can be visualized as being the product of
(i)~a circle, corresponding to the gauge degree of freedom $\theta$, and
(ii)~a sphere, corresponding to the angular variables $\{\gamma, \beta\}$)
(see Fig.~\ref{fig:orderparameter_visual}).
The parameters $(\theta,\gamma)$ and $(\theta+\pi,\gamma+\pi)$ give identical values of the order parameter [see Eq.~(\ref{eqn:op_def})], and therefore correspond to physically identical configurations.

As we aim to discuss states having time-reversal symmetry breaking, we have assumed the Ginzburg-Landau parameter $\sigma$ (defined in Eq.~\ref{eqn:TensorParameters}) to be positive as, for sufficiently weak $\text{SO}(2)_z$ symmetry-breaking, this favors states in which $\beta = \pm \pi/2$. These states are related by time-reversal symmetry, and can be visualized as lying at the poles of the order-parameter sphere (see Fig.~\ref{fig:orderparameter_visual}).
In these states, the order parameter takes the form
\begin{equation}
    {\bm \op}|_{\beta=\pm\pi/2}
    =
    \opmag e^{i(\theta\mp\gamma)} \frac{1}{\sqrt{2}}
    \left( \begin{array}{c} 1 \\ \pm i \\ \end{array} \right).
\end{equation}
If, across the entire system, the state had chirality $\beta=\pi/2$, only a single, position dependent, phase field would be required to describe low-energy excitations away from equilibrium (and, similarly, if the state had only chirality $\beta=-\pi/2$). On the other hand, to describe low-energy excitations featuring both chiralities, as well as the \lq\lq domain walls\rq\rq\ between them (through which $\beta$ changes between $\pm\pi/2$), a pair of position-dependent phase fields, $\theta(\posvec)$
and $\gamma(\posvec)$, is required.  It will often be convenient to exchange these fields for the pair
\begin{equation}
\label{eqn:thetaalpha}
\theta^{\pm}({\bm \posvec}):=\theta({\bm \posvec})\mp\gamma({\bm \posvec}).
\end{equation}
From Eq.~(\ref{eqn:fpot}) we can see that within a domain of maximally positive (or maximally negative) $\beta$, the free energy does not depend on $\theta^+$ (or $\theta^-$), and this remains true even after weak $\rm{SO}(2)_z$ symmetry-breaking terms are included in $f_\text{P}$.
Consequently, the subspace of energy-degenerate homogeneous equilibrium states is disconnected, being composed, topologically, of two circles, which exchange under time-reversal, i.e. $\degensubspace= S^1 \times \{+,-\}$
(see~Fig.~\ref{fig:orderparameter_visual}). This order-parameter space combines two of the most familiar order-parameter spaces: the $S^1$ of conventional superconductivity/superfluidity, and the $\{+,-\}$ of Ising magnetism.

In general, to analyze the topological structure of order parameters, we consider their homotopy groups $\pi_n(\degensubspace)$ associated with $\degensubspace$.  For the specific case of $\degensubspace=S^1 \times \{+,-\}$, since each connected piece is isomorphic to $S^1$ the first homotopy group $\pi_1(\degensubspace) \cong \mathbf{Z}$. This implies that a domain of a given chirality can exhibit vortex singularities, as, e.g., in the case of conventional superconductivity.  As the space $\degensubspace$ is topologically disconnected, the zeroth homotopy group is also nontrivial, i.e., $\pi_0(\degensubspace) \cong \mathbf{Z}_2$; this implies the possibility of domain walls, which separate regions of opposing chirality.  (Domain walls are common features of systems in which the order parameter is discrete as in the Ising case.)\thinspace\
The $\mathbf{Z}_2$ value of $\pi_0$ indicates that domain walls annihilate one another.
We remind the reader that order parameters for which $\pi_n(\degensubspace)$ is nontrivial support topological defects of co-dimension $n+1$.  Thus, in the effectively two-dimensional (real) space that we are considering, vortices  points and domain walls are lines.

The domain walls and vortices determine the {\it qualitative} structure of order-parameter field configurations; e.g., vortices in a domain of positive or negative chirality correspond to topological singularities in $\theta^+(\posvec)$ or $\theta^-(\posvec)$.  In particular, when there are a total of $N^\pm$ vortices at positions
$\{{\bm R}^\pm_\nu\}_{\nu}^{N^\pm}$ having vorticities $\{q^\pm_\nu\}_{\nu}^{N^\pm}$ interior to the positive- (or negative-) chirality domain, the singularities of $\theta^\pm$ can be characterized by the local {\it vortex density (scaler) fields} $\rho_\textrm{v}^+$ and $\rho_\textrm{v}^-$, which are defined via
\begin{equation}
\label{eqn:vrho}
    2 \pi \rho_\textrm{v}^\pm({\bm \posvec}):=
    E_{ab} \nabla_a \nabla_b \theta^\pm({\bm \posvec})
    = 2 \pi \sum_{\nu=1}^{N^\pm} q^\pm_\nu \delta ({\bm \posvec} - {\bm R}^{\pm}_\nu).
\end{equation}
Here and elsewhere in this paper, the Dirac delta functions $\delta$ are are softened on an appropriate lengthscale; for vortices it is the vortex core diameter.

Domain walls also have implications, but for the qualitative structure of $\beta({\bm \posvec})$.  In two spatial dimensions, domain walls are lines, and a collection of $N$ then can be characterized by specifying their trajectories $\{{\bm R}_n(s)\}_{n=1}^{N}$ as functions of an arclength parameter $-s_{n}\le s\le s_{n}$.  By requiring, in addition, that the unit vector normal to the domain wall $\hat{n}_a(s)$,
which is related to the domain wall trajectory via
\begin{equation}
\label{eqn:domain_wall_normal}
\hat{n}_a(s)= (\cos \phi(s), \sin \phi(s) )_a = - E_{ab}\,\partial_{s} R_b(s),
\end{equation}
point from the negative towards the positive chiral domain, the sense of the vector tangent to the domain wall, $\partial_s {\bm R}(s)$, is determined.  It is natural to associate the locations of the domain walls with the zeros of the field $\beta(\posvec)$; for a given set of domain walls, the equilibrium form of $\beta(\posvec)$ interpolates smoothly---with a solitonic form whose thickness is then the domain wall width---between regions in which it is essentially uniform and equal either to $\pi/2$ or to $-\pi/2$.  Such structure can be characterized via a {\it domain-wall density (vector) field} ${\bm \dwrhodown}$, defined via
\begin{equation}
\label{eqn:dwrho}
    {\bm \dwrhodown}({\bm \posvec}) :=
   \frac{1}{2} {\bm \nabla} \sin \beta_\text{dw}({\bm \posvec})
    \approx \sum_{n=1}^N  \int_{-s_n}^{s_n}\!\!\! ds \, \hat{{\bm n}}(s) \, \delta(\posvec-\textbf{R}_n(s)).
\end{equation}
Here, the delta function is softened on the lengthscale of the domain-wall width.  We shall make use of the vortex and domain wall densities given in Eqs.~(\ref{eqn:vrho},\ref{eqn:dwrho}) in Sec.~\ref{sec:effective_theory} in the construction of an effective free energy for the the topological variables.

\begin{figure}
\includegraphics[width= .2\textwidth]{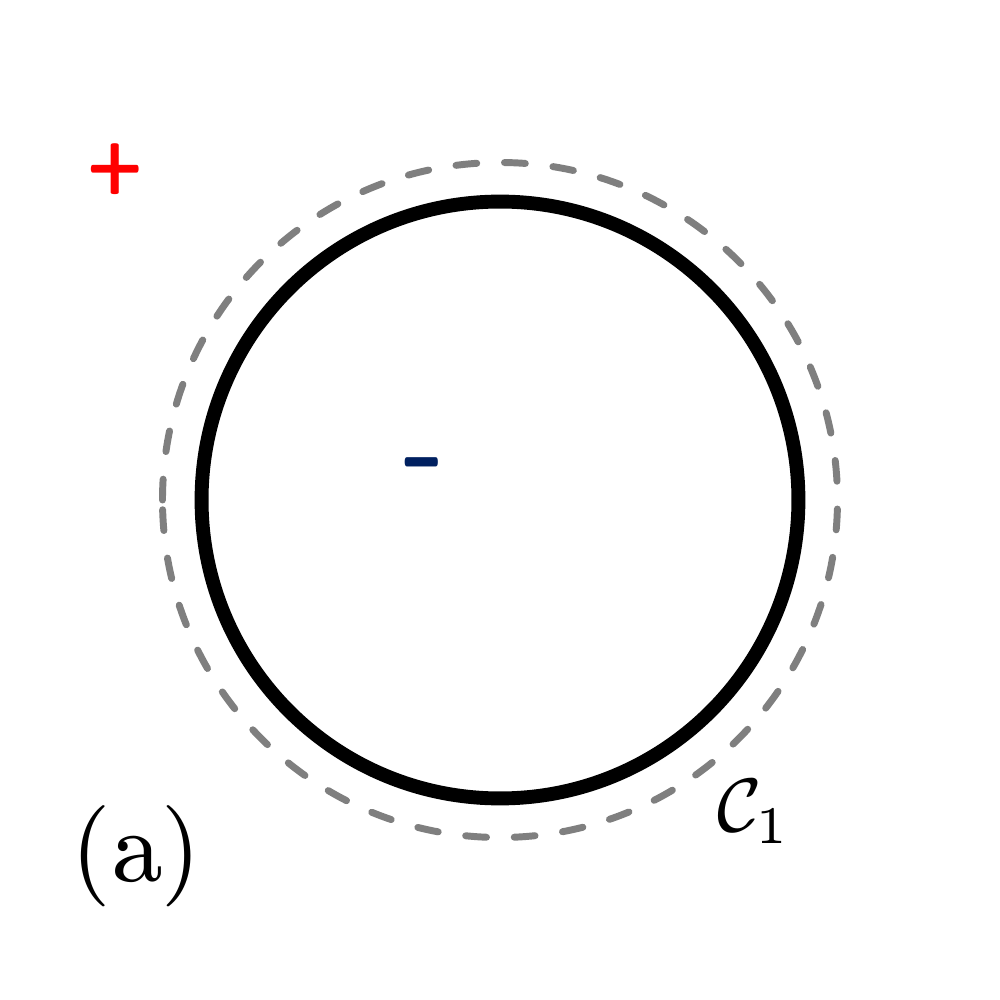}
\includegraphics[width= .2\textwidth]{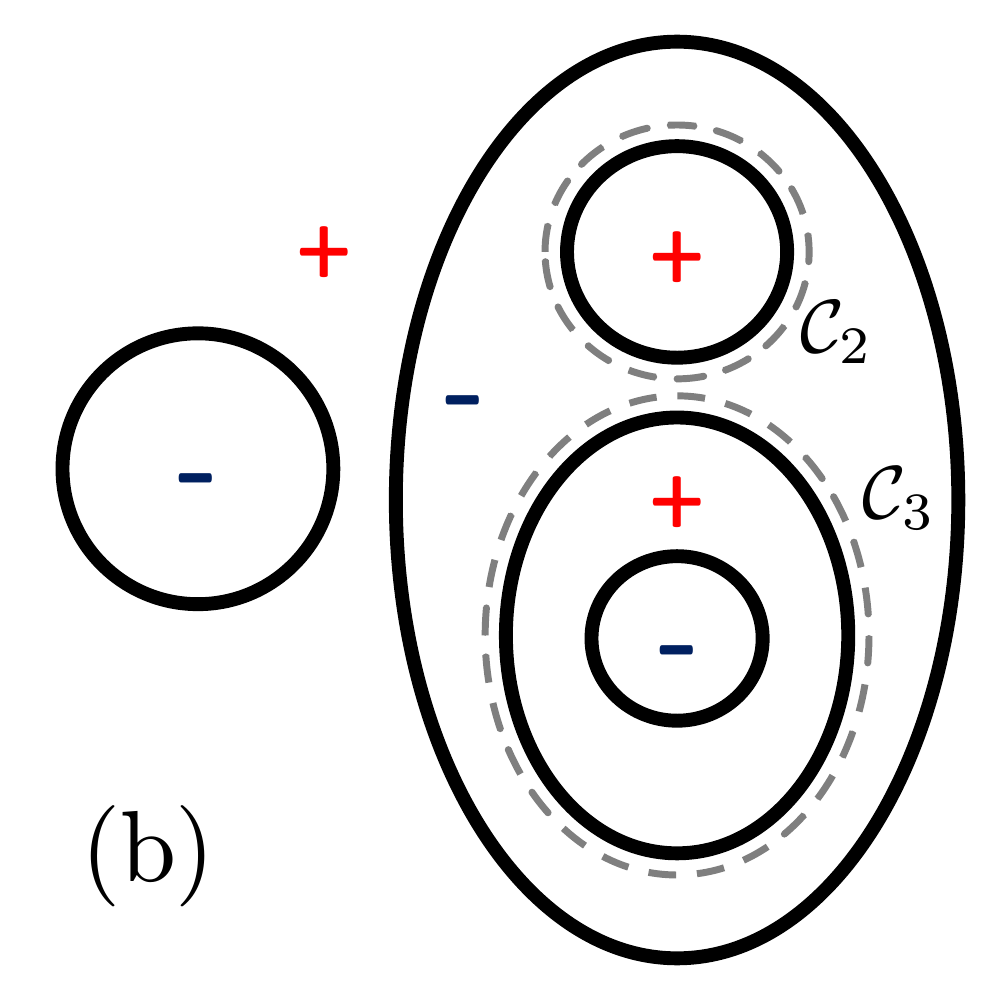}
\caption{(a)~A large region of positive chirality having an internal-island domain of negative chirality.  As the positive-chirality region is multiply connected, the winding of $\theta^+$ around the contour ${\cal C}_1$ is an independent topological variable.
(b)~Six, connected chiral regions (including the exterior, positive-chirality region).  For every multiply connected chiral region, there is an independent topological variable associated with each non-contractible loop. As an example, the multiply connected negative-chirality region has two independent, non-contractible loops ${\cal C}_2$ and ${\cal C}_3$.  Each of these loops is associated with an independent winding number for  $\theta^-$.  (color online)
\label{fig:island_domains}}
\end{figure}
It should be recognized, however, that these densities do not, by themselves, fully specify the topological structure of the order-parameter field.  To see this, note, e.g., that a single island chiral domain affects the topology of the surrounding domain by rendering it multiply connected.  Thus, to fully specify the topological structure of the $\theta^+(\posvec)$ and $\theta^-(\posvec)$ fields---in addition to specifying the location and vorticity of the individual vortices that lie within the respective domains---the global winding of of $\theta^+(\posvec)$ or $\theta^-(\posvec)$ must be specified around a loop that encircles each island (see Fig.~\ref{fig:island_domains}a).  To generalize to chiral domain structures that involve islands within islands, we note that to fully specify the topological structure of each positive (negative) connected chiral domain region, one must specify the winding of the $\theta^+$ ($\theta^-$) field around each independent non-contractible loop of that region (see Fig.~\ref{fig:island_domains}b)~\footnote{We note that for weak coupling and specular reflection, the condition at the sample boundary requires $\beta=0$ (see, e.g., Ref.~\cite{Sigrist91}), and thus, for a finite sample, all regions can be considered to be surrounded by domain-wall loops. However, the coupling between $\dgfGamma$ and the normal direction of either the surface or the domain wall is generically different, which can result in distinct equilibrium orientations of $\dgfGamma$ for each of these cases (see, e.g. ~\cite{Matsumoto99}).}.

In the remainder of this section we explain the connection between the $\gamma(\posvec)$ field on a single domain-wall loop surrounding an island and the determination of the global freedom to wind possessed by the multiply-connected region exterior to the island.  We also examine various situations involving individual chiral-domain islands, chosen to illustrate the physical roles played by the $\gamma(\posvec)$ field.  Before doing this, we remark that on any domain-wall line ${\bm R}(s)$ (i.e., a locus of points on which $\beta = 0$), the order parameter takes the form
\begin{equation}
\label{eqn:gamma_at_dw}
    {\bm \eta}\vert_{\beta=0} = \opmag e^{i \theta}
                        \left(
                          \begin{array}{c}
                            \cos \gamma \\
                            \sin \gamma \\
                          \end{array}
                        \right).
\end{equation}
Thus we see that the field $\gamma$ evaluated along a domain wall trajectory  ${\bm R}(s)$, defines an angular variable $\dgfGamma(s):=\gamma\big({\bm R}(s)\big)$ for each value of the arclength parameter $s$. We furthermore see that the function $\dgfGamma(s)$ determines the structure of the order parameter along the domain wall-line, specifically via the planar vector $(\cos\dgfGamma(s) ,\sin \dgfGamma(s))$.

\begin{figure}
\includegraphics[width= .2\textwidth]{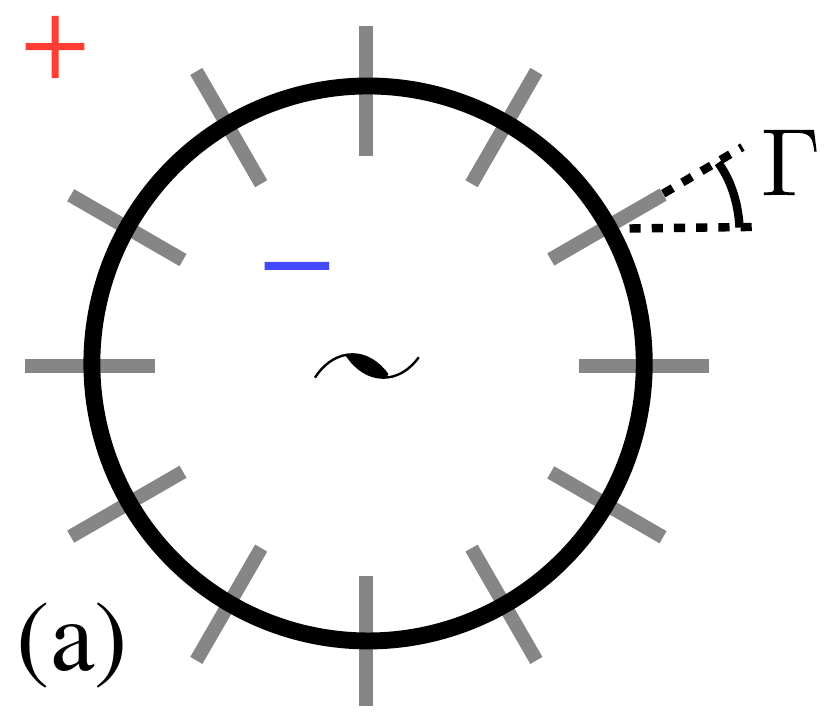}
\includegraphics[width= .2\textwidth]{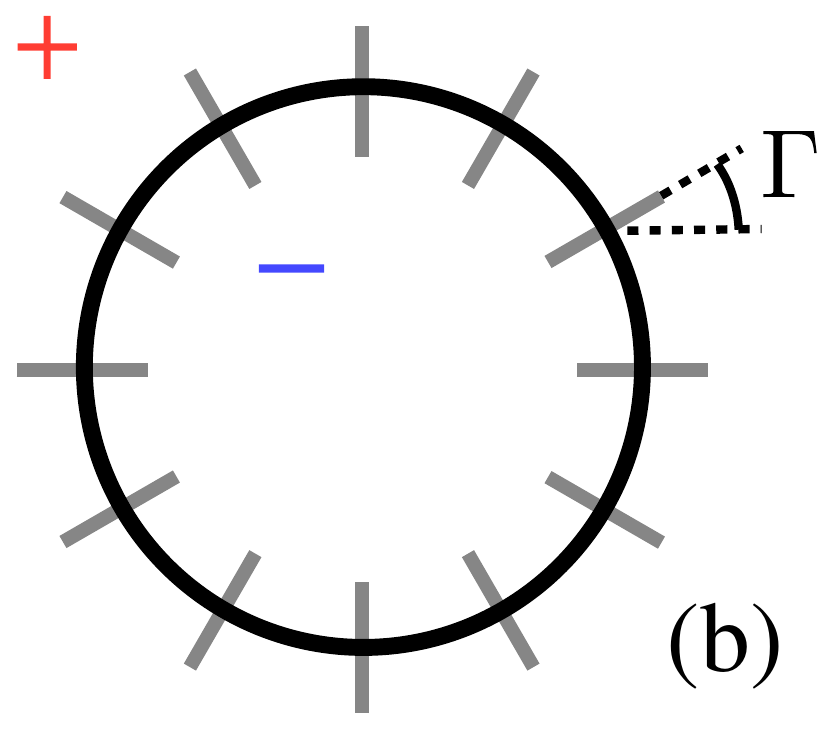}
\includegraphics[width= .2\textwidth]{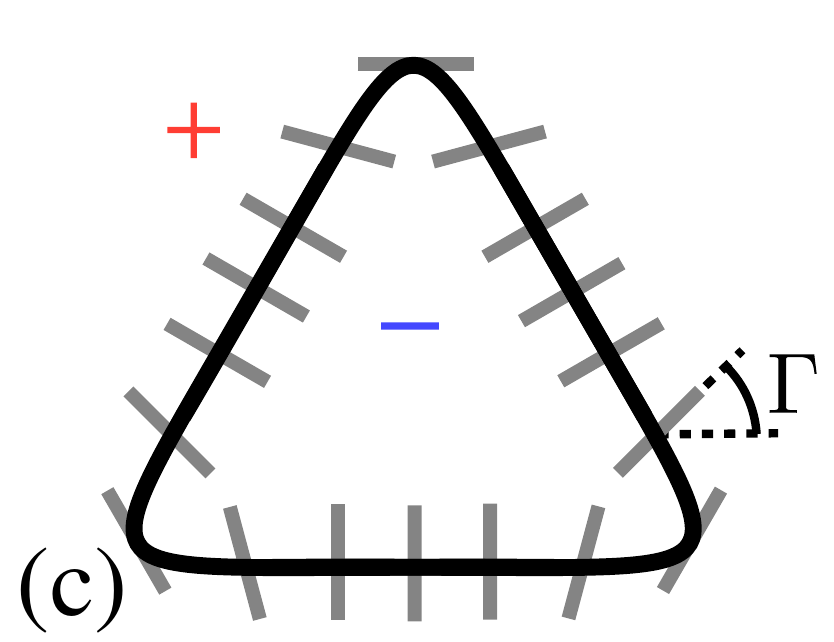}
\caption{Three types of domain wall loops (black lines). The gray line segments intersecting the domain wall indicate the local direction of the planar vector $(\cos\dgfGamma(s),\sin\dgfGamma(s))$ (i.e., the strongest-pairing direction).  Each loop is characterized by the three winding numbers $(n_+,n_-,n_\gamma)$ [see Eq.~(\ref{eqn:dw_loop_winding_numbersTG},\ref{eqn:dw_loop_winding_numbersPM})]: (a)~$(-1,1,1)$; (b)~$(-2,0,1)$; and (c)~$(1,0,-1/2)$. These domain wall loops are topologically equivalent to vortices with the following properties (a)~singly quantized and rotationally invariant; (b)~coreless and rotationally invariant; and (c)~singly quantized and coreless. (color online)
\label{fig:chiral_island}}
\end{figure}
To illustrate the physical role played by $\dgfGamma(s)$, we begin by considering the special case of a simply-connected chiral domain, bordered by a domain wall, and thus interior to a larger surrounding region of opposing chirality.  Two topological numbers $(n_\theta,n_\gamma)$, defined as follows, characterize the winding of the order parameter along paths that lie near to the domain wall:
\begin{equation}
\label{eqn:dw_loop_winding_numbersTG}
\begin{split}
    2 \pi n_\theta &:= \oint_\textrm{dw} d \theta \, ,
    \\
    2 \pi n_\gamma &:= \oint_\textrm{dw} d \dgfGamma \,,
\end{split}
\end{equation}
where ``dw'' indicated that the line integrals are evaluated along the domain wall, and the orientation of these integrals is taken to be counterclockwise.  From Eq.~(\ref{eqn:thetaalpha}), we see that $(n_\theta,n_\gamma)$ provide the same information as the two topological numbers $( n_+,n_- )$, defined via
\begin{equation}
\label{eqn:dw_loop_winding_numbersPM}
    2 \pi n_\pm := \oint_\textrm{dw} d \theta^\pm \,;
\end{equation}
specifically, $n_{\pm}=n_\theta\mp n_\gamma$.  We note that $n_\theta$ and $n_\gamma$ are either both integral or both half-integral [so that $\bm\eta$ is single-valued; see Eq.~(\ref{eqn:gamma_at_dw})], and thus that $n_+$ and $n_-$ are both integral.

For the sake of definiteness, we consider a domain of negative chirality that constitutes a simply-connected island within a larger, positive chiral domain. The positive domain is then rendered multiply connected; see Figs.~\ref{fig:island_domains} and~\ref{fig:chiral_island}. Each of the winding numbers $(n_+,n_-,n_\gamma)$ corresponds to a distinct physical property:
\begin{itemize}
\item The winding number $n_+$ of the exterior (positive) domain essentially determines, via $\Phi_{\rm tot}=\Phi_0 n_+$, the total flux $\Phi_{\rm tot}$ through an area that extends beyond the region bounded by the domain wall by a few penetration depths.
\item The winding number $n_-$ of the interior (negative) domain is the net number of vortices in the interior domain; if $n_- = 0$ then the domain-wall loop can be coreless, i.e., there is no topological requirement that there exist locations where $\opmag$ vanish.
\item Whether or not the winding number $n_\gamma$ is unity determines whether or not the the superconducting order can be rotationally invariant around a circular domain wall.
\end{itemize}
Importantly, as we previously noted in this section, by specifying the vorticial content in the interior and exterior domains, the interior winding number {\it is} uniquely determined, whereas the exterior winding number {\it is not}. However, if in addition to the vorticial content, the value of $n_\gamma$ is specified, the winding of the exterior domain is also determined. In the particular case under consideration, $n_+ = n_- -2 n_\gamma$.

Thus, $n_\gamma$ plays a dual role, determining both the total flux $\Phi_{\rm tot}$, via its influence on $n_+$ as well as whether or not the superconducting order can be rotationally invariant.

We pause to make two remarks concerning issues of energetics.  First, as a domain wall has finite energy-cost per unit length, to reduce its energy a domain wall loop may shrink in size. When viewed on a lengthscale much larger than the domain size, a small domain wall loop appears topologically equivalent to a vortex~\cite{Izyumov90}, and thus provides an alternative description of the various vortex structures that can occur in superconductors with broken time-reversal symmetry (see, e.g., Refs.~\cite{Tokuyasu90,Heeb99,Sauls09}). Second, in the limit in which the free energy is rotationally invariant and $\kappa$ is large, energy considerations prefer domain-wall loops that are singly quantized (i.e., contain flux $\Phi_0$), rotationally invariant, and coreless.  However, owing to the linear relationship between $n_+$, $n_-$, and $n_\gamma$, these preferences cannot all be simultaneously satisfied.  Compromise order-parameter configurations result from this frustration; we show in Fig.~\ref{fig:chiral_island} examples of the three types of vortices that satisfy two of the three preferences. Which particular type of vortex is preferred, energetically, will depend on the details of the parameters in the Ginzburg-Landau theory.

To illustrate this frustration and the dual physical role played by $n_\gamma$, we now consider two of the three small domain-wall loops that are favored energetically in the extreme London limit. In this limit, energetic considerations allow only coreless vortices, e.g., $n_- = 0$, and this implies that $\Phi_{\rm tot} = - 2 \Phi_0 n_\gamma$. Thus we see that, in the extreme London limit, if we also impose rotational invariance, namely $n_\gamma = 1$, we effect the magnetic properties of the vortex, requiring the vortex to be doubly quantized (i.e., contain $2 \Phi_0$ of flux) and fixing the sense of the magnetic flux. Conversely, if we fix the vortex to be singly quantized then, in the extreme London limit, the vortex would not be rotationally invariant. This interplay between the rotational and magnetic structure, perpetrated by the dual nature of the $\gamma$ field, underpins the central results of this work.

So far, we have established that, when taken together with vorticial content, $n_\gamma$ fixes the {\it overall\/} winding of the order parameter along a non-contractible loop within a multiply-connected chiral domain.
However, to describe the {\it local\/} structure of a domain wall [see Eq.~(\ref{eqn:gamma_at_dw})], it is necessary to specify the local value of $\gamma$ [viz.~$\dgfGamma(s)$] along the domain wall.  As we shall see in Secs.~\ref{sec:effective_theory} and \ref{sec:bending_domain_wall-second}, the local behavior of $\dgfGamma(s)$ also plays a role in determining the flux that penetrates through the domain wall locally~\cite{Sigrist89, Sigrist99}.  Thus, in order to develop a \emph{local} description of the superconductor, the natural degree of freedom to use---for specifying the additional topological structure afforded by the presence of multiply-connected regions---is $\dgfGamma(s)$ rather than $n_\gamma$.

In the following section, by starting with the Ginzburg-Landau free energy, we construct an effective local free energy in terms of the topological variables. Specifically, we show that, in addition to the vortex density and domain wall density, the free energy depends on a third topological variables, viz., the value of $\dgfGamma(s)$ along domain walls.

\section{Effective free energy in terms of topological descriptors and the Extended London limit}
\label{sec:effective_theory}

The aim of this section is to begin with the description of the superconducting system in terms of a Ginzburg-Landau free-energy functional dependent on the order-parameter field, and to derive from it a reduced description in terms of the vortex and domain-wall densities and $\dgfGamma(s)$ along domain walls.  In this reduced description, the focus is on the dependence of the free energy on the topological variables [i.e., the locations of the vortices and domain wall, as well as $\dgfGamma(s)$].  The presumption is that the degrees of freedom associated with exciting the order parameter {\it around} the state of minimum free energy within a fixed topological sector [defined via the locations of the vortices and domain walls and $\dgfGamma(s)$] have been eliminated, either by integrating them out or by setting them to their stationary values.  For a numerical implementation of the latter procedure applied to superconductivity in Sr$_2$RuO$_4$, see Ref.~\cite{Ichioka05}.  In the present paper, our aim is to proceed analytically, a task that is eased by our working in a particular limiting regime, an elaboration of the standard London limit that we term the \lq\lq extended London limit\rlap.\rq\rq\thinspace\  We remind the reader that the standard London limit amounts to assuming that the Ginzburg-Landau parameter $\kappa$ tends to infinity, which enforces the condition $\opmag=1$.  To pass to the extended London limit we make the additional assumption that the width of domain walls, which is controlled by the parameter $\Ldw$, tends to zero.  In this limit, the domain wall becomes vanishingly thin, compared with the penetration depth.

We begin with the Ginzburg-Landau free energy, Eq.~(\ref{eqn:dimnesionlessGL}), and first pass to the standard London limit.  From Eq.~(\ref{eqn:fpot}), we see that in this limit the order-parameter amplitude $\opmag$ is energetically prohibited from departing from unity; inserting the corresponding form of the order parameter [i.e., Eq.~(\ref{eqn:op_def}) but with $\opmag=1$] into Eq.~(\ref{eqn:dimnesionlessGL}), making the definition
$(\Delta\Theta)_{a i} := (\nabla_a \theta - A_a , \nabla_a \gamma , \nabla_a \beta )_i$,
and dropping constant terms arising from the potential terms, we arrive at the unconventional superconducting free energy $F_{\rm L}$ corresponding to the London free energy for conventional superconductivity, i.e.,
\begin{eqnarray}
    \label{eqn:FLondon}
    &&F_{\rm L}[\theta,\gamma,\beta,{\bm A},{\bm H}] =
    \nonumber\\
    &&\qquad\, \int d^{2}r \bigg\{
    \frac{1}{2} K_{a b c d}\, \Xi_{b i d j}\, (\Delta \Theta)_{a i}\, (\Delta \Theta)_{c j}
    \nonumber\\
    &&\qquad \qquad+
    \frac{1}{8 \Ldw^2} \cos^2 \beta
    +\frac{1}{2} \vert ({\bm \nabla} \times {\bm A}) - {\bm H} \vert^2
    \bigg\},
    \\
    &&\,\,\,\,\,\, \xi_{a i} : = R^\gamma_{ab}
    ( i I^{\phantom{1}}_{bc} ,-E^{\phantom{1}}_{b c}, i M^1_{bc})_i (\cos \beta, i \sin \beta)_c,
    \nonumber\\
    &&\Xi_{a i b j} : = \frac{1}{2}(\xi^*_{ai} \xi_{bj} + \xi^*_{bj} \xi_{ai}),
    \nonumber
\end{eqnarray}
where repeated indices $i,j,\ldots$ are summed from 1 to 3.  Because in this free energy, the coefficient $\Xi$ is contracted with a tensor that is symmetric under time reversal, we have adopted a form for $\Xi$ that is manifestly symmetric under time reversal.  In general, the supercurrent density ${\bm J}(\posvec)$ is given by
$-\delta F_{\rm sc}/\delta{\bm A}(\posvec)$
and, continuing within the London limit, we see that it has the form
\begin{subequations}
\label{eqn:currentsLondon}
\begin{eqnarray}
    J_a &&= g_{ab}(\nabla_b\theta - A_b) + {\jun}_a,
    \\
     {\jun}_a&&:=
    -\sin \beta \, \nabla_a \gamma
    \nonumber\\
    &&\phantom{:=} +\frac{1}{4}(2 \mu \cos \beta \, \matE_{ab} + \tau {\cal M}^{\gamma}_{ab} )\nabla_b \beta,
    \\
    {\cal M}^\gamma_{ab}&&:= R^\gamma_{ac} R^\gamma_{bd} M^1_{cd} =
        \left(
        \begin{array}{cc} -\sin 2 \gamma & \cos 2 \gamma\\
        \phantom{-}\cos 2 \gamma  & \sin 2 \gamma  \\ \end{array}
        \right)_{ab},
        \\
        g_{ab}&&:=\matI_{ab} + \frac{1}{2}\tau \cos \beta {\cal M}^{\gamma-\pi/4}_{ab}\,.
\end{eqnarray}
\end{subequations}
Note the occurrence of the unconventional contribution ${\bm\jun}$ to the supercurrent, which includes currents that are localized near domain walls~\cite{Volovik85}.  This contribution is manifestly odd under time reversal (which is evident because each term is odd in $\beta$).

We now proceed to take the {\it extended} London limit, in which domain walls are controlled to be thin compared with the penetration depth.  We begin by noting that the term arising from $f_{\rm P}$ that remains in the free-energy density in the London limit is $\cos^2(\beta)/8 \Ldw^2$, and that this term contributes positively for any value of of $\beta\ne\pm \pi/2$.  In particular, for a domain wall, across which $\beta$ varies from $\pi/2$ to $-\pi/2$, the balancing, in equilibrium, of this potential term against contributions to the free energy that result from gradients in $\beta$ would produce a spatial configuration in which  $\beta$ changes from $\pi/2$ to $-\pi/2$ over a lengthscale (i.e., the domain-wall width) proportional to $\Ldw$.  Thus, in the limit $\Ldw\to 0$, the widths of domain walls are controlled to be arbitrarily small, compared with the penetration depth (which, we remind the reader, we have chosen to set the unit for lengths).

This extension of the London limit results in useful simplifications.  First, as the domain walls  are arbitrarily thin, regions in which $\beta$ is uniform and equal to $\pm \pi/2$  dominate, areally.  Thus, terms proportional to $\cos \beta$ or $\sin \beta$ become $0$ or ${\rm sgn} \beta$ respectively.  [Note that $ {\rm sgn}\beta$ is the unit step function, taking the values $1$ (or $-1$) for regions of positive (negative) chirality i.e., $\beta>0$ (or $\beta<0$)]. As an explicit example, the term in the superfluid density tensor ${\bm g}$ proportional to $\cos\beta$ can be neglected in the extended London limit, and thus we may make the replacement ${\bm g} \to {\bm I}$.  Physically, this means that, even in the presence of domain walls, the in-plane Meissner response is isotropic.

A second useful simplification that arises in the extended London limit is that it enables us to express contributions to the free energy and supercurrent involving gradients of $\beta$ in terms the domain-wall density ${\bm \rho}_{\rm dw}$, defined in Eq.~(\ref{eqn:dwrho}).  Using Eq.~(\ref{eqn:currentsLondon}) we can thus, e.g., write
\begin{subequations}
\begin{eqnarray}
    {\bm J} &&= {\bm \nabla} \theta - {\bm A} + {\bm \jun},
    \\
    {\bm \jun} &&:=
    -{\rm sgn} \beta \, {\bm \nabla} \gamma
    +\big( \mu {\bm \matE} + \frac{\pi}{4}\tau{\bm {\cal M}}^{\dgfGamma(s)} \big) \cdot {\bm \rho}_{\rm dw}\,.
\end{eqnarray}
\end{subequations}
For the sake of compactness, here and elsewhere we use the notation ${{\bm {\cal M}}^{\dgfGamma(s)} \cdot {\bm \rho}_{\rm dw}}$ as shorthand for
$\int ds\,{\cal M}^{\dgfGamma(s)} \cdot {\bm n}(s)\,\delta({\bm \posvec} - {\bm R}(s))$.

Having discussed how, in the extended London limit, the spatial variation of $\beta$ is fully incorporated via the locations of the domain walls $\{ {\bm R}_n(s)\}$, we continue with our goal of constructing an effective free energy by eliminating all degrees of freedom associated with excitations of the order parameter around the state of minimum free energy within a fixed topological sector.  With this in mind, our next step is to eliminate the non-topological variations in the $\theta$ field.

As the superconducting order may possess vortices, $\theta$ is not, in general, a single-valued function of position, and therefore it may exhibit singular behavior (i.e., at the cores of vortices).  Our initial strategy for eliminating the non-topological variations of $\theta$ is to decompose it into two components: $\theta = \theta_{\rm sm} + \theta_{\rm v}$, where $\theta_{\rm sm}$ is a smooth, single-valued part, and $\theta_{\rm v}$ is the part that accounts for any vortex singularities.  This separation is not unique, but we shall see, after eliminating $\theta_{\rm sm}$ from the free energy by setting it equal to its stationary value $\bar{\theta}_{\rm sm}$, that the resulting free energy is---for any fixed choice of topological variables, such as vortex positions and strengths---independent of any particular choice of decomposition.  To implement this elimination of $\theta_{\rm sm}$ we need only consider the terms in the free-energy density associated with the kinetic energy of the supercurrents (i.e., associated with $J^2$), as other terms do not depend on $\theta$; in the extended London limit the free energy $F_{J}$ constructed from these terms is given by
\begin{equation}
\label{eqn:Fcurrents}
    F_{J} = \int d^{2}r\,\frac{1}{2}
    \left\vert
    {\bm \nabla} \theta_{\rm sm} + {\bm \nabla} \theta_{\rm v} - {\bm A} + {\bm \jun} \right\vert^{2}.
\end{equation}
Stationarity of this expression with respect to $\theta_{\rm sm}$ reads
\begin{equation}
-\nabla^2 \theta_{\rm sm}=
{\bm \nabla} \cdot ( {\bm \nabla} \theta_{\rm v} - {\bm A} + {\bm \jun} ),
\end{equation}
and, by using the Green function for the Laplace operator in two dimensions
[i.e., $G({\bm \posvec})=-\frac{1}{2\pi}\ln\vert {\bm\posvec}\vert$, obeying
$-\nabla^{2}G({\bm\posvec})=\delta({\bm\posvec})$],
the stationary solution $\bar{\theta}_{\rm sm}$ can be expressed as
\begin{equation}
    {\bar \theta}_{\rm sm}({\bm \posvec}')=
    \int d^{2}r\,G({\posvec}'-\posvec){\bm\nabla}\cdot
    \big({\bm \nabla}\theta_{\rm v}-{\bm A}+{\bm\jun}\big)({\bm \posvec}).
\end{equation}
By inserting ${\bar \theta}_{\rm sm}$ into Eq.~(\ref{eqn:Fcurrents}), we arrive at the following form for the free energy:
\begin{subequations}
\label{eqn:TransverseCurrents}
\begin{eqnarray}
    &&F_{J}
    =\frac{1}{2}\int d^{2}r\,\big\vert{\bm J}^{\rm T}\big\vert^{2},
    \\
    &&J^{\rm T}_a ({\bm\posvec})
    :=\int d^{2}r^{\prime}\,
    \Big(
    I_{ab}\,\delta({\bm\posvec}-{\bm\posvec}')
    - \nabla_{a}G({\bm\posvec}-{\bm\posvec}')\,\nabla'_{b}
    \Big)
   \nonumber \\
    &&\qquad\qquad\qquad\qquad
    \times\Big(\nabla_b \theta_{\rm v} - A_b + \jun_b\Big)
    \\
    &&=\int d^{2}r^{\prime}\,
    E_{ab}\nabla_{b}\,G({\bm\posvec}-{\bm\posvec}')\,
    E_{cd}\nabla'_{c}
    \Big(\nabla'_d\,\theta_{\rm v}-A_{d}+\jun_{d}\Big),
    \nonumber
\end{eqnarray}
\end{subequations}
where we have used the elementary result $E_{ab}E_{cd}=I_{ac}I_{bd}-I_{ad}I_{bc}$ and the defining equation obeyed by $G$.
The procedure of minimizing $F_{J}$ with respect to $\theta_{\rm sm}$ can be described, physically, as compensating for any source of longitudinal currents (i.e., current-flows that build up at some location) or, equivalently, as a projection on to the subspace of transverse currents.
This construction brings to the fore the vorticial content of the transverse supercurrent, which arises both from vortices and domain walls. Specifically, one can identify the vorticity $W$ via
\begin{equation}
W=E_{ab}\nabla_{a}\big(J_{b}+A_{b}\big)
 =E_{ab}\nabla_{a}\big(\nabla_{b}\theta_{\rm v}+\jun_{b}\big).
\end{equation}
Owing to the unconventional contribution to the supercurrent ${\bm \jun}$, the vorticity $\vort$ in unconventional superconductivity in the extended London limit comprises both a vortex term $\vort_{\rm v}$, which is common also to conventional superconductivity and is proportional to the total vortex density $\vrho$, and a domain-wall term $\vort_{\rm dw}$, which is proportional to the domain-wall density ${\bm\dwrhodown}$:
\begin{subequations}
\begin{eqnarray}
    \label{eqn:vorticity-top}
    \vort_{\phantom{t}} &=& \vort_\text{v} + \vort_\text{dw},\\
    \nonumber
    \vort_\text{v}  &:=& 2 \pi\bigg(
    \frac{1}{2}\left(1+\text{sgn}\beta\right) \vrho^+({\bm \posvec})\\
    \label{eqn:vorticity-mid}
    &&\qquad+
    \frac{1}{2}\left(1-\text{sgn}\beta\right) \vrho^-({\bm \posvec})
    \bigg),
    \\
    \label{eqn:vorticity-bottom}
    \vort_\text{dw} &:=&
    \big(
    f(s)\,\textbf{n}(s)
    +{\bm d}(s)\cdot{\bm \nabla}\big)\cdot{\bm\dwrhodown},
\end{eqnarray}
\end{subequations}
where
\begin{subequations}
\label{eqn:dipole_monopole_vorticity}
\begin{eqnarray}
    f(s)
    &:=&
    -2 \, \partial_s \dgfGamma(s),
    \label{eqn:monopole_vorticity}\\
    {\bm d}(s)
    &:=&
    -\mu \,
    {\bm \matI}
    +\frac{\pi}{4} \tau \,
    {\bm R}^{2(\phi(s) - \dgfGamma(s) )}.
    \label{eqn:dipole_vorticity}
\end{eqnarray}
\end{subequations}
Several points are worth noting here. First, $\vort_\text{v}$ is a weighted sum of the vortex densities in the chiral domains, which makes evident the fact that only those singularities of $\theta^+$ ($\theta^-$) that are located in the positive-chirality (negative-chirality) domain are associated with local vorticity.  Second, via Eq.~(\ref{eqn:vorticity-bottom}), we see that the domain wall vorticity $\vort_{\rm d}$ can be expressed as a sum of two contributions: a ``monopole'' contribution of strength $f$, which determines the net magnetic flux penetrating the superconductor; and a ``dipole'' contribution of strength ${\bm d}$, which is generated by currents that flow along domain wall cores but do not create net flux through the superconductor.  Third, within this extended London limit, the monopole and dipole contributions are expressible in terms of the topological degrees of freedom $\dgfGamma(s)$ and $\phi(s)$.  We remind the reader that $\phi(s)$ is determined from the trajectory of a domain wall ${\bm R}(s)$ via Eq.~(\ref{eqn:domain_wall_normal}).

The final step in deriving the reduced free energy is to eliminate the vector potential ${\bm A}$.
Although it is possible to proceed directly, using Eq.~(\ref{eqn:TransverseCurrents}) (see Appendix~\ref{sec:apdx_conventional_eff}), the fact that the current ${\bm J}^{\rm T}$ in Eq.~(\ref{eqn:TransverseCurrents}) is determined via a nonlocal expression makes it more efficient to apply an alternative,  \lq dual\rq\ approach, which uses a Hubbard-Stratonovich transformation of the nonlocal kernel via an auxiliary field $\Hdual$; see, e.g., Ref.~\cite{Polyakov87,Zee03}.
The resulting, dual expression for the free energy $F_{J}$ is thus given by
\begin{equation}
    F_{J}[\Hdual]= \int d^{2}r\,
    \left\{ -\frac{1}{2} \vert {\bm \nabla} \Hdual\vert^2 + \Hdual ( W - B) \right\}.
    \label{eqn:dualfree}
\end{equation}
Under the constraint that it be evaluated at the stationary value of $\Hdual$, this form for $F_{J}$ has the same value as the one given in Eq.~(\ref{eqn:TransverseCurrents}). We note, in passing, that the dual free energy $F_{J}[\Hdual]$ depends explicitly on the local value of the perpendicular magnetic field $B$ ($= E_{ab} \nabla_a A_b$). Thus, the full expression for the free energy in the extended London limit, which also includes the magnetic field energy $\int d^{2}r\,\frac{1}{2}(B-H)^2$, depends on $B$ locally.  This locality renders simple the task of identifying the stationary value of $B$.  Eliminating $B$ by setting it equal to its stationary value we arrive at the following form for the extended London limit of the free energy:
\begin{equation}
\label{eqn:FullDual}
    F_{\rm EL}\!=\!\!\int\!\!d^{2}r
    \left\{\!\!-\frac{1}{2}\Hdual(-\nabla^2 \!+\!1)\Hdual
    \!+\!\Hdual( W \!-\! H)\!+\!\!\fdwcore\!\right\}.
\end{equation}
In this form, the first two terms, which together account for the kinetic energy of the supercurrent and the magnetic field energy, have the virtue of being local and determined via $\vort$ (i.e., the vorticity of the supercurrent).  The remaining contributions to the free energy given by Eq.~(\ref{eqn:FLondon}) are accounted for via $\fdwcore$, which is associated with the core energy of the domain walls and are negligibly small in regions lying beyond a distance of a few wall widths $\Ldw$ from a domain wall.  An explicit expression for $\fdwcore$ in terms of the fields $\gamma$ and $\beta$ is given in Appendix~\ref{sec:straight_dw}~\footnote{Here and elsewhere in this paper, we take into account the core energies of domain walls but not the core energies of vortices.  Our justification for doing this is that, in the standard London limit, the energy cost of a vortex core is negligibly small, compared with the kinetic energy of the supercurrents and magnetic fields, whereas the core energy of a domain wall is not.}.\thinspace\
Thus, in the neighborhood of the extended London limit, in which $\Ldw$ becomes small (but remains non-zero), the domain-wall energy $\int d^{2}r\,\fdwcore$ can be expressed in terms of an energy  per unit domain-wall length $E_{\rm core}$, which depends locally upon on $\dgfGamma(s)$ (i.e., $\gamma$ evaluated on the domain wall) together with the shape of the domain wall [(e.g., via $\phi(s)$)], along with their arclength derivatives:
\begin{equation}
       \int d^{2}r\,\fdwcore=
       \sum_{n}\int_{-s_n}^{s_n}ds\,E_{\rm core}\left(\dgfGamma_n(s),\ldots;\phi_n(s),\ldots\right).
\end{equation}

We are now in the position to complete our derivation of the reduced free energy $F_{{\rm EL}}$ in the extended London limit, reduced in the sense that it depends only on the external applied magnetic field and the topological variables via the vorticity $\vort$ and and domain-wall core energy density $\fdwcore$.  Upon eliminating $\Hdual$ from Eq.~(\ref{eqn:FullDual}), $F_{{\rm EL}}$ becomes
\begin{eqnarray}
     \label{eqn:Feff}
     &&F_{\rm EL}=\int d^{2}r\,\fdwcore
     \\
     &&\!\!\!\!\!+\!\int{\frac{d^2r\,d^2r'}{4\pi}}
     \big(\vort(\textbf{r})\! -\! H(\textbf{r})\big)
     K_0(|\textbf{r}\!-\!{\textbf{r}}'|)
     \big(\vort({\textbf{r}}') \!- \!H({\textbf{r}}')\big),
\nonumber
\end{eqnarray}
where $K_0$ is a modified Bessel function of the second kind.  A virtue of the formulation that we have employed is that it enables the efficient calculation of the magnetic response of the superconductor in the extended London limit, via the thermodynamic relation
\begin{equation}
\label{eqn:M_calc}
    M_{\rm EL}({\bm \posvec}) = -\frac{\delta F_{\rm EL}}{\delta H({\bm r})} =
    \int \frac{d^2r'}{2 \pi}
    K_0(|\textbf{r} - \textbf{r}'|)
    \big(\vort({\textbf{r}}') - H({\textbf{r}}')\big).
\end{equation}
One can also use the Amp\`ere-Maxwell law to determine the spatial distribution of equilibrium supercurrents in the this limit, which gives ${\bm J}_{\rm EL} = {\bm \matE} \cdot {\bm \nabla} M_{\rm EL}$.

As an initial illustration of this approach, we consider a straight domain wall, lying along the $y$-axis in infinite, three-dimensional superconductor.  We assume that there is no applied magnetic field, i.e., $H=0$.  We further assume that the superconducting state is of positive (negative) chirality for $x<0$ ($x>0$), so that by the convention defined by Eq.~(\ref{eqn:domain_wall_normal}) we have ${\bm R}(s) = s\hat{\bm y}$.  As we show in the Appendix~\ref{sec:straight_dw}, a variational analysis, based on an assumed form for the behavior of $\beta$ transverse to a translationally invariant domain wall, suggests that the equilibrium value of $\dgfGamma$ is $\phi$.  Assuming this to be case, we then find, from Eq.~(\ref{eqn:dipole_monopole_vorticity}), that the domain-wall vorticity has no monopole part [i.e., $f(s)=0$] but does have a dipole part, which is given by
${\bm d}=\big((\pi\tau/4)-\mu\big){\bm \matI}$~\footnote{In the case that the equilibrium value of $\dgfGamma - \phi$ is $\pi/2$, the dipole part of the vorticity is given by ${\bm d}=\big(-(\pi\tau/4)-\mu\big){\bm \matI}$.  Thus, as is shown in the referring paragraph, the part of the domain-wall current proportional to $\tau$ can flow in either direction, for a given pattern of chirality, depending on the value of $\dgfGamma - \phi$.}. Then, from Eq.~(\ref{eqn:M_calc}), we find that magnetization  and current densities vary with the spatial distance $x$ from the domain wall as follows:
\begin{subequations}
\begin{eqnarray}
\label{eqn:M_EL}
    M_{\rm EL}(x)
    &=& -\frac{1}{2}\big((\mu - (\pi \tau/4 )\big)\,\text{sgn}(x)\,e^{-\vert x \vert},\\
\label{eqn:J_EL}
    {\bm J}_{\rm EL}(x)
    &=&\big(\mu - (\pi \coeffmirror/4)\big)
       \big(\delta(x) - e^{-\vert x \vert}/2\big)\,\hat{\rm y}.
\end{eqnarray}
\end{subequations}
As, for this  domain-wall configuration, the monopole contribution $f(s)$ to the domain-wall vorticity is zero, the net magnetic flux (per unit length of domain wall) [e.g. the magnetic flux (per unit length of domain wall) integrated transversally] vanishes. The jump discontinuity in $M(x)$ at $x=0$ results from a supercurrent that flows along the domain-wall core.  For the case of Sr$_2$RuO$_4$, we can use Eq.~(\ref{eqn:J_EL}) to estimate the magnitude of this current.
In SI units the dimensionful current density ${\bm J}'$ is given in terms of its dimensionless counterpart ${\bm J}$ via
${\bm J}' = (2\pi\lambda f_0/\Phi_0){\bm J}$.
Then, using ${\bm J}$ to compute the current passing through a narrow window bracketing the domain wall, we arrive at the following expression for the dimensionful domain-wall current $I'$ per Ru-O layer:
\begin{subequations}
\begin{eqnarray}
    I' &=&\frac{2\pi\lambda^{2} f_0}{\Phi_0}
    \Delta z\int_{0^{-}}^{0^{+}}\!\!\!
    dx\,\left(\mu - (\pi\tau/4)\right)\delta(x)
    \\
    &=&\frac{2\pi\lambda^{2} f_0}{\Phi_0}
    \Delta z\,\left(\mu - (\pi\tau/4)\right),
\end{eqnarray}
\end{subequations}
where $\Delta z$ is the thickness of an Ru-O layer.
To find the numerical value of this current in Amps, we note that in SI units $f_0$ can be expressed in terms of the thermodynamic critical field $H_c$ as $2\kappa^{-2}\mu_0 H^2_c$.
Using the parameter values appropriate for Sr$_2$RuO$_4$~\cite{Mackenzie03}, i.e.,
$\mu_0 H_c = 0.023\,{\rm T}$,
$\lambda(0) = 0.15\,\mu\rm{m}$,
$\kappa=2.3$, and
$\Delta z = 1.2\,{\rm nm}$,
we arrive at the following estimate for the current:
$\big(\mu-(\pi \tau/4)\big)\times{1.3}\times 10^{-5}\,{\rm A}$
per Ru-O layer flowing along a domain-wall core.
For this result to match previously made theoretical estimates (see Ref.~\cite{Matsumoto99,Kwon03}), one would need to have the material perameters obey $\big(\mu - (\pi \coeffmirror/4)\big)\approx 1$.

In the next section, we extend our discussion to cope with situations lying beyond straight domain walls, thus allowing the domain walls to have bends. As part of this discussion, we employ the reduced description of the superconductor in the extended London limit derived in the present section to show that: (i)~a net magnetic flux penetrates the superconductor near bends; and (ii)~this flux is generically a {\it nonintegral multiple} of the superconducting flux quantum $\Phi_0$.

\section{Magnetic flux in the vicinity of a bend in a domain wall}
\label{sec:bending_domain_wall}

In this section we derive the central result of this work, viz., that a bend in a domain wall is accompanied by a nonintegral amount of magnetic flux that penetrates the superconductor near the bend; the amount---which we term the {\it bend flux\/}---depends on the geometry of the bend.  In the limit in which the in-plane crystalline anisotropy is negligible (i.e., the isotropic limit), the bend flux is proportional to the angle through which the domain wall bends.

We derive the bend flux via two approaches. In the first, we analyze a bending domain wall via the effective theory of the topological variables, developed in Sec.~\ref{sec:effective_theory}.  We then consider an alternative derivation, which, in the isotropic limit, yields the bend flux quite generally, without reliance on the assumption of either the standard or the extended London limit, or even on the validity of the Ginzburg-Landau expansion of the free energy.  We end this section by considering modifications of the isotropic-limit bend flux result that would arise in settings of other pairing symmetries and/or tetragonal or hexagonal departures from the limit of crystalline isotropy.

\subsection{Comparison with an a spatially extended Josephson junction}
\label{sec:bending_domain_wall-first}
Before establishing the existence of bend flux, we give a discussion of the the essential differences between, on the one hand, a system comprising a domain wall and the superconducting regions of opposing chirality separated by it, and, on the other hand, a system of a spatially extended Josephson junction and two regions of conventional superconductivity coupled by it.  For the extended Josephson-junction system it is possible to define a variable analogous to the domain wall variable $\dgfGamma(s)$, i.e., the local value $\dgfGamma_\text{J}(s) := (-\theta_1(s) + \theta_2(s))/2$ of (half of the) the difference between the phases $\theta_{1}(s)$ and $\theta_{2}(s)$ of the superconducting regions that lie on either side of the junction.  The important distinction between $\dgfGamma$ and $\dgfGamma_{\rm J}$ is that whereas $\dgfGamma_{\textrm J}$ transforms trivially under in-plane rotations, $\dgfGamma$ transforms nontrivially.

\begin{figure*}
\includegraphics[width = \textwidth]{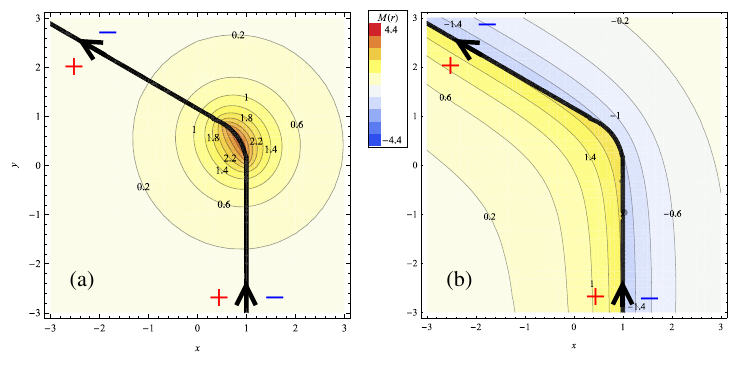}
\caption{(a) ``Dipole'' and (b) ``monopole'' contributions to the magnetic field associated with a bending domain wall (oriented black line) for a bend angle of $\pi/3$, considered in the extended London limit [see Eqs.~(\ref{eqn:dipole_monopole_vorticity}) and~(\ref{eqn:M_calc})]. The $z$-axis magnetic field $M({\bm \posvec })$ is plotted as a function of position (color scale and contour lines).  For this example, the Ginzburg-Landau parameters are taken to have the values $\mu=0.2$ and $\tau=1$.  The penetration depth defines the unit lengthscale.  The chirality is positive to the left of the domain wall and negative to the right of it, so that, via Eq.~(\ref{eqn:domain_wall_normal}), we see that the domain wall has the indicated orientation. Only the monopole contribution produces a net magnetic flux through the system. (color online)
\label{fig:flux_dw_bend}}
\end{figure*}
This observation has important implications, if we compare the local energy of a domain wall $E_{\rm dw}$ with the Josephson energy $E_{\rm J}$ of a extended Josephson junction.  In particular, for the extended Josephson junction, $E_{\rm J}$ is a periodic function of $\dgfGamma_{\rm J}$ alone.  For the domain wall system, on the other hand, in order to preserve the ${\rm SO}(2)_z$ invariance of the free energy, the local energy-density of the domain wall must be a periodic function of the difference $\dgfGamma-\phi$ [in which $\phi$ is determined by the local direction of the domain wall normal; see Eq.~(\ref{eqn:domain_wall_normal})].
Thus, because they have distinct values of $\phi$, two segments of straight domain wall separated by a bend will generically have distinct equilibrium values of $\dgfGamma$.  This stands in contrast with the case of the spatially extended Josephson junction with a bend, the equilibrium value of $\dgfGamma_{\rm J}$ being independent of position along the junction.  As we shall now see, the bend flux originates in this variation of the equilibrium value of $\dgfGamma$ on either side of a bend.

\subsection{Bend flux in terms of topological variables}
\label{sec:bending_domain_wall-second}
We now turn to the derivation of the bend flux within the special context of the effective theory for topological variables, developed in the previous section.  Part of the utility of this effective theory is that it allows for an efficient calculation of the magnetic response of the superconductor, given a configuration of the topological variables, viz., the position and strength of the vortices, the positions of the domain-wall lines, and the value of $\dgfGamma(s)$ along each such line.  Thus, our approach will be to consider a specified configuration of topological variables {\it without\/} vortices but with a single, fixed domain wall having a bend and a specified form for $\dgfGamma(s)$ along it, and then to employ Eq.~(\ref{eqn:M_calc}) in order to determine the corresponding magnetization density.

We define the position of the domain wall using three line-segments: an arc of $\bendangle$ radians and unit radius of curvature, and two straight segments that continue tangentially from each of the end-points of the arc (see Fig.~\ref{fig:flux_dw_bend}).  Given this particular geometry, we say that the resulting domain wall has a bend angle of $\bendangle$ in it. Our next assumption concerns the form of $\dgfGamma(s)$. In Appendix~\ref{sec:straight_dw} we give a variational analysis that suggests that, for a straight domain wall, the equilibrium value of $\dgfGamma$ is $\phi$. To generalize to the situation in which the domain wall bends, we assume that $\dgfGamma$ follows the local direction of the domain wall \lq adiabatically\rlap,\rq\ i.e., $\dgfGamma(s) = \phi(s)$.  In this case, because $\partial_s \dgfGamma(s)$ is not everywhere zero a monopole contribution to the domain-wall vorticity arises [see Eqs.~(\ref{eqn:vorticity-bottom}) and~(\ref{eqn:monopole_vorticity})], in addition to the dipole contribution.  Figure~\ref{fig:flux_dw_bend} shows both the monopole and dipole contributions to the magnetic field, evaluated using Eq.~(\ref{eqn:M_calc}).

Next, we determine the total flux $\Phivarbend$ associated with the bent domain wall furnished by this variational calculation. To do this, we integrate the total magnetic field through a large circular disc $\Omega$ centered at the vertex formed by the extrapolation of the straight-line segments, so that the straight-line segments lie radial to the disc.  In the limit that the disc radius is much larger than the penetration depth, we find that the dipole contribution to $\Phivarbend$ tends to zero, whereas the monopole contribution is nonzero, tending to the following total flux:
\begin{equation}
    \label{eqn:eff flux through bend}
    \Phivarbend = \int_\Omega d^{2}r\,M({\bm x})= 2\int d\dgfGamma(s)=2\bendangle,
\end{equation}
i.e., the net flux is simply given by twice the bend angle, regardless of how $\dgfGamma(s)$ interpolates between its limiting values far from the bend.  In particular, for case shown in Fig.~\ref{fig:flux_dw_bend} (i.e., for $\bendangle = \pi/3$) the bend flux is $2 \pi /3$, i.e., the dimensionful value is $\Phi_0/3$, which is a nonintegral multiple of the flux quantum.

\subsection{General analysis for the bend flux}
\label{sec:bending_domain_wall-third}
\begin{figure}
\includegraphics[width= .4\textwidth]{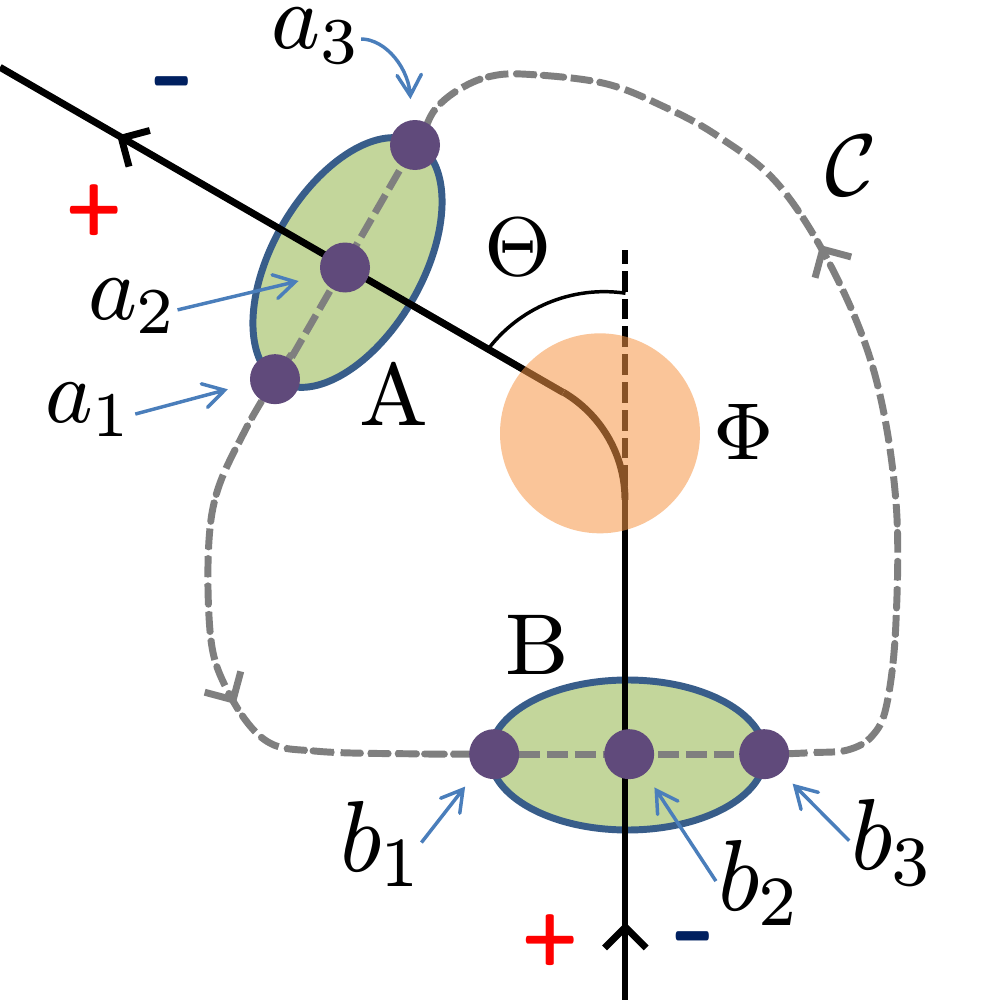}
\caption{A bent domain wall (oriented black line) separating two domains of opposite chirality. En route to deriving Eq.~(\ref{eqn:Bend_Angle_Flux}), which expresses the the flux $\Phi$ through the area bounded by the contour $\cal{C}$ in terms of the bend angle $\bendangle$, the circulation of the unconventional superfluid velocity ${\bm \sfv}$ [see Eq.(\ref{eqn:sfvelocity})] around $\cal{C}$ is shown to be zero. The contour $\cal{C}$ is assumed to be many penetration depths away from the region where the domain wall bends. (color online)
\label{fig:general_case}}
\end{figure}
In the remainder of this section we consider a more general context in which the existence of bend flux can be demonstrated. In particular, we need not employ the extended London limit, nor assume that the superconductor is in the Ginzburg-Landau regime.  Rather, the central assumption is that the superconducting order has the following essential feature: in regions in which the chirality is maximal, local ${\rm SO}(2)_z$ rotations of the superconducting order can equally well be accomplished via ${\rm U}(1)$ gauge transformations, so that the local transformation
$(\alpha, -{\rm sgn}( \beta) \, \alpha ) \in {\rm U}(1) \times {\rm SO}(2)_z$
acts {\it trivially\/} on the superconducting order parameter.  If this assumption holds then, provided the {\it amplitude\/} of the superconducting order is spatially homogeneous away from the domain wall, the unconventional superfluid velocity ${\bm \sfv}$, defined via
\begin{equation}
\label{eqn:sfvelocity}
{\bm \sfv} := {\bm \nabla}\theta-\text{sgn}(\beta)\,{\bm \nabla}\gamma-{\bm A},
\end{equation}
tends to zero within a maximally chiral region, as a result of the Meissner effect.

In deriving the bend flux we use the following construction to define the geometry of the domain wall. We consider a single domain wall that is fixed to pass through three points: the origin $O$, and two other points $P$ and $Q$; then, we fix the angle $\angle POQ = \pi + \bendangle$ where $\bendangle$ is  bend angle of the domain wall.  We take the orientation of the domain wall, as defined by Eq.~(\ref{eqn:domain_wall_normal}), to run from $P$ to $Q$, and we let the points $P$ and $Q$ tend to infinity.  The goal of the calculation is then to determine the net magnetic flux penetrating the superconductor in the vicinity of the domain wall bend.  The key quantity that we use is the circulation of ${\bm \sfv}$ around a closed contour $\cal{C}$ encircling the domain wall bend at a distance of many penetration depths (see Fig.~\ref{fig:general_case}).  Care is needed in selecting the contour $\cal{C}$ because, in equilibrium, even for zero applied magnetic field, a domain wall may not be translationally invariant~\footnote{For example, as noted in Ref.~\cite{Volovik85}, in the limit of large domain-wall currents, vortices may be stabilized along the domain wall.  Such vortices would then spoil the translational invariance of the domain wall.  Although we know of no experimental evidence for such an effect, there is evidence from Josephson-junction tunneling experiments on Sr$_2$RuO$_4$~\cite{Kidwingira06} for the absence of translational invariance (i.e., for the presence of multiple domains) along sample boundaries, which can be regarded in certain respects as analogous to domain walls.}.  However, as the underlying free energy is local [as is manifest in Eq.~(\ref{eqn:FullDual})] and translationally and rotationally invariant; it is always possible to choose two geometrically congruent regions, A and B, each straddling the domain wall
but located on opposite sides of the domain-wall bend, in which the equilibrium spatial configurations of the superconducting order in the regions (A and B) are related to one another via a rigid rotation and translation (see Fig.~\ref{fig:general_case}).  Once a pair of such regions has been identified, we choose the contour $\cal{C}$ to cross the regions (and hence the domain wall) on locally identical paths $b_1 \rightarrow b_3$ and $a_1 \rightarrow a_3$ (i.e., on paths that are related by the same rotation and translation as the regions are).  As a result, the following equality between line integrals holds:
\begin{equation}
\label{eqn:vcancel}
    \int_{a_1 \rightarrow a_3}d{\bm \posvec}\cdot {\bm \sfv}
    =
    \int_{b_1 \rightarrow b_3}d{\bm \posvec}\cdot {\bm \sfv}.
\end{equation}
By using this result, and observing that ${\bm \sfv} = {\bm 0}$ away from the domain wall, we see that the circulation of ${\bm \sfv}$ around the closed contour $\cal{C}$ is zero, i.e.,
\begin{equation}
\label{eqn:zero-circ}
    \oint_{\cal{C}} d{\bm \posvec}\cdot {\bm \sfv}
    =
    0.
\end{equation}

The next step in the derivation is to consider the contour $\cal{C}^+$ ($\cal{C}^-$), which begins at the point $a_2$ ($b_2$) and follows $\cal{C}$ through the positive- (negative-)chirality domain to the point $b_2$ ($a_2$). The line-integrals of $({\bm\sfv}+{\bm A})$ along $\cal{C}^+$ and $\cal{C}^-$ respectively measure the change in the phase of the order parameter in the positive (negative) region from $a_2$ to $b_2$ (and from $b_2$ to $a_2$). Thus, again using the linear relation $\gamma = (-\theta^+ + \theta^-)/2$ [i.e., Eq.~(\ref{eqn:thetaalpha})], we see that the change in $\Gamma$ from the point $b_2$ to the point $a_2$
[i.e., $\Delta \dgfGamma:= \dgfGamma(a_2)-\dgfGamma(b_2)$],
is given by the following formula:
\begin{equation}
\label{eqn:delta-gamma}
    \Delta \dgfGamma=
    \frac{1}{2}\int_{\cal{C}^+} d\theta^+
    +
    \frac{1}{2}\int_{\cal{C}^-} d\theta^-
    =
    \frac{1}{2}\oint_{\cal{C}} d{\bm \posvec} \cdot ({\bm\sfv} +{\bm A}).
\end{equation}
We now examine in more detail the equilibrium value of $\Delta \dgfGamma$.  As discussed in Sec.~\ref{sec:bending_domain_wall-first}, as a consequence of the rotational invariance of the underlying free energy, the energy (per unit arclength)  $E_{\rm dw}$ of the domain wall must be a periodic function of the combination $\dgfGamma(s)-\phi(s)$, in which $\phi(s)$ continues to characterize the local direction normal to the domain wall. Furthermore, as---up to a global phase---the configurations having $\dgfGamma$ and $\dgfGamma+\pi$ are equivalent, the dependence of $E_{\rm dw}$ on $\dgfGamma(s)-\phi(s)$ has period $\pi$.  Importantly, we make the following additional assumption, viz., that the dependence of $E_{\rm dw}$ on $\dgfGamma(s)-\phi(s)$ has a {\it single minimum} per period.

We now observe that, by construction, region~A is rotated by an angle $\Theta$ relative to B (using the convention that positive rotations are measured counter-clockwise, relative to the domain-wall orientation) as a result $\Delta \phi := \phi(a_1)-\phi_(a_2) = \bendangle$.
Thus, with these assumptions the equilibrium value of $\Delta \dgfGamma$ is equal to the bend angle $\Theta$, up to an integer multiple of $\pi$, i.e.,
\begin{equation}
\label{eqn:change-alpha}
    \Delta \dgfGamma = \Theta + n \pi.
\end{equation}
Combining Eqs.~(\ref{eqn:zero-circ},\ref{eqn:delta-gamma}, \ref{eqn:change-alpha}),
and defining $\Phibend$ to be the bend flux (i.e., net flux through the surface defined by the contour $\cal{C}$) we arrive at the result that
\begin{equation}
\label{eqn:Bend_Angle_Flux}
    \Phibend = \big((\Theta/\pi)+ n\big)\,\Phi_0.
\end{equation}
Because the bend flux, in the rotationally invariant limit, can evidently be an arbitrary fraction of the flux quantum, this result is a manifestation of the general result that broken time-reversal invariance allows for nonquantized amounts of flux to penetrate a superconductor, as predicted on general grounds in Refs.~\cite{Volovik84,Geshkenbein87,Sigrist89,Sigrist95,Sigrist99,Volovik00}.  Moreover, because $\dgfGamma(s)$ need not stay locked, relative to the local domain wall orientation (e.g, at the bend), or owing to the presence of vortices in either or both of the chiral domains, it makes sense that $\Phibend$ be determined only modulo $\Phi_0$ \footnote{As a particular case of Eq.~(\ref{eqn:Bend_Angle_Flux}), one can consider a straight domain wall. Then, Eqs.~(\ref{eqn:change-alpha}) and (\ref{eqn:Bend_Angle_Flux}) imply that the topologically stable, localized solitons in $\dgfGamma(s)$ along a domain wall would obey $\Delta \dgfGamma = \pi$, and that each is associated with a flux $\Phi_0$.  However, if the dependence of $E_{\rm dw}$ on $\dgfGamma(s)-\phi(s)$ should have multiple minima per $\pi$ period, then there could be topologically stable solitons in $\dgfGamma(s)$ along a straight domain wall, each having $\Delta \dgfGamma\neq \pi$ and connected with nonquantized amounts of flux (see Refs.~\cite{Sigrist89,Sigrist99,Bouhon10}). In this case, in addition to the bend flux of Eq.~(\ref{eqn:Bend_Angle_Flux}), the flux associated with a bent domain wall may have a further contribution.}.

\subsection{Bend fluxes for other pairing and crystalline symmetries}
\label{sec:other-symms}
In this paper, we have assumed that the superconducting order transforms as one particular representation of ${\rm SO}(2)_z$.  We now obtain the generalization of the formula for the bend flux, Eq.~(\ref{eqn:Bend_Angle_Flux}), that remains valid for arbitrary irreducible representations, which can be indexed in terms of an integer $m$ (see, e.g., Ref.~\cite{Landau77}). For brevity's sake, we refer to the $m=1$ case as p-wave (which is the case focused on in this paper), and the $m=2$ case as d-wave. En route to generalizing Eq.~(\ref{eqn:Bend_Angle_Flux}) to arbitrary $m$, we assume that transformations of the form $(m \, \alpha, -{\rm sgn} (\beta) \, \alpha ) \in {\rm U}(1) \times {\rm SO}(2)_z$ act trivially on a uniform, maximally chiral phase.  Under this assumption, and repeating the line of argument given in Sec.~\ref{sec:bending_domain_wall-third}, {\it mutatis mutandis\/}, the bend flux formula becomes
\begin{equation}
\Phi_{\rm bend,m} = \big((m \Theta/\pi)+ n\big)\,\Phi_0.
\end{equation}

Another version of Eq.~(\ref{eqn:Bend_Angle_Flux}) results when we address the setting of tetragonal $D_{4h}$ symmetry (which is, of course, discrete).  In this case, the argument given in Sec.~\ref{sec:bending_domain_wall-third} leading to Eq.~(\ref{eqn:Bend_Angle_Flux}) holds only for $\bendangle=\pm\pi/2$, for which the minimum net flux through the domain wall bend in the p-wave case is $\Phi_0/2$; this is distinct from a conventional vortex, for which the net flux is $\Phi_0$.  In contrast, for the d-wave case and $\bendangle = \pm \pi/2$, a net flux of $n\Phi_0$ (with $n$ integral) penetrates the bend.  As a last observation, we note that for p-wave pairing and $D_{6h}$ symmetry and a domain-wall bend angle of $\pi/3$, the smallest {\it positive\/} net flux accompanying the bend is $\Phi_0/3$, whereas the smallest {\it negative\/} net flux accompanying it is $-2\Phi_0/3$.

Now that we have established that in, various settings, one anticipates that a bent domain wall is accompanied by nonintegral net magnetic flux, we shall, in the next section, explore a range of experimental situations in which such effects might be observable.

\section{Experimental implications}
\label{sec:Experiments}
\begin{figure}
\includegraphics[width= .3\textwidth]{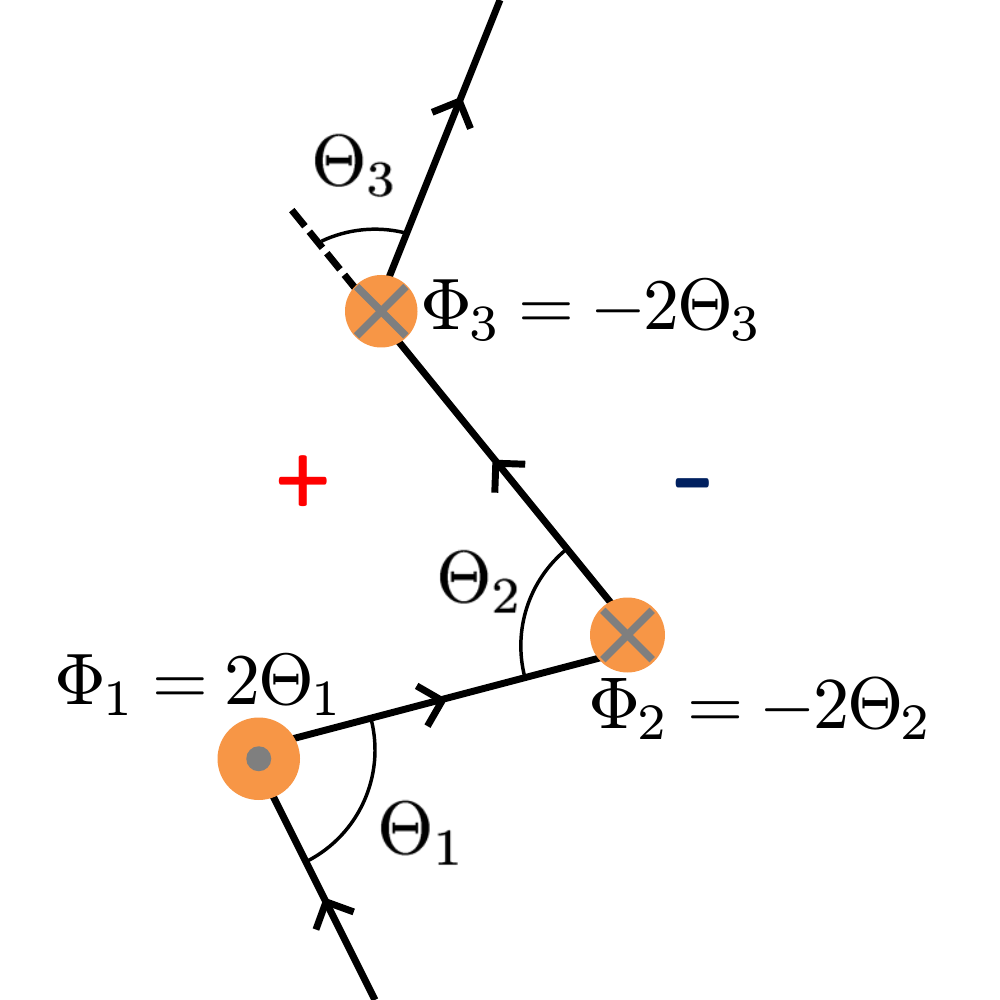}
\caption{Schematic depiction of a chiral domain wall running along an ab-face of a superconductor and pinned to various locations.  The wall is indicated by the oriented black line [the orientation is defined by Eq.~(\ref{eqn:domain_wall_normal})]. We assume that deviations from ${\rm SO}(2)_z$ symmetry are sufficiently small that the arrangement of the pinning sites determines the path of the domain wall.  Orange dots denote pinning sites.  Near them, the domain wall bends and flux penetrates the superconductor.  The specified positive angles $\{\Theta_i\}$ express the geometry of the bends, as indicated.  The bend fluxes are then determined via Eq.~(\ref{eqn:Bend_Angle_Flux}).  For each bend, we have chosen the value of $n$ in Eq.~(\ref{eqn:Bend_Angle_Flux}) to give the corresponding bend fluxes $\{ \Phi_i \}$ the smallest possible magnitudes. The orientation of the flux accompanying each bend is indicated via a dot (up) or a cross (down). (color online)
\label{fig:weak_pinning}}
\end{figure}

\begin{figure}
\includegraphics[width= .3\textwidth]{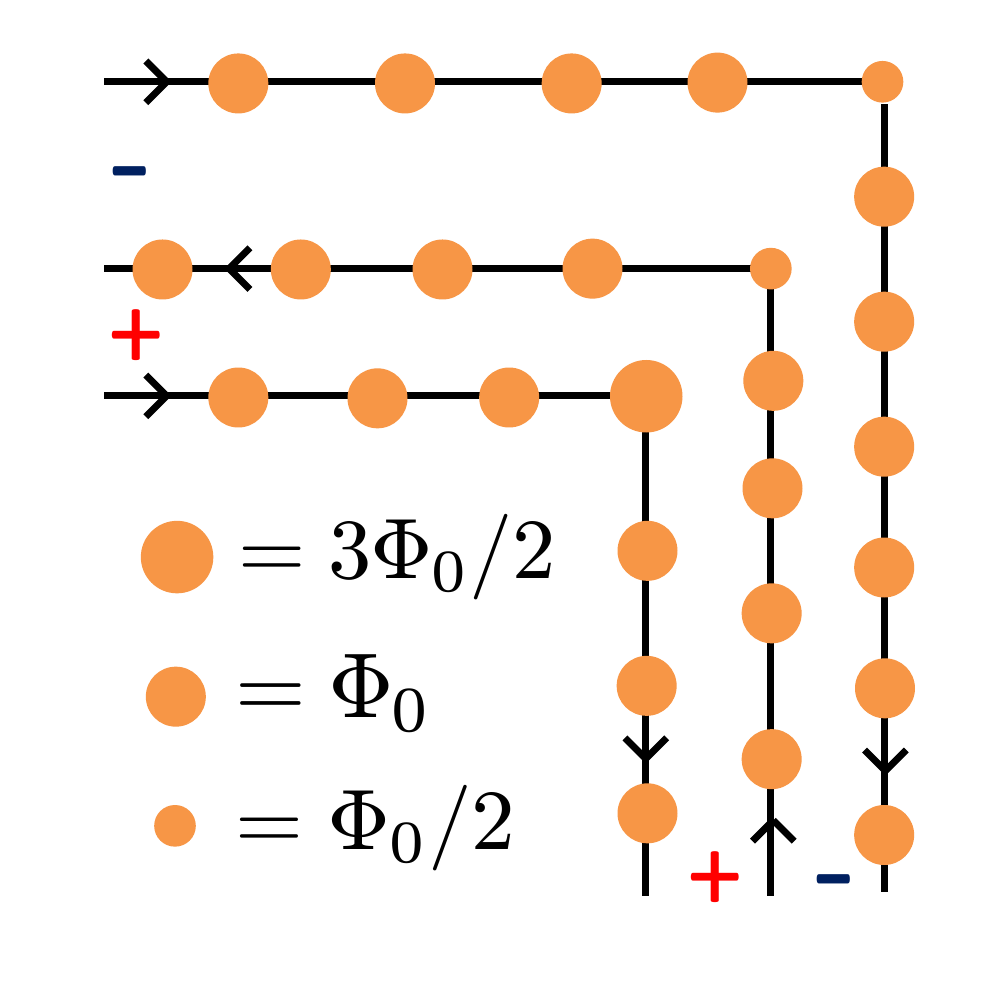}
\caption{Schematic depiction of an array of domain walls (oriented black lines), each intersecting an ab-face of a superconductor. $D_{4h}$ deviations from ${\rm SO}(2)_z$ symmetry are assumed to be large enough to pin domain walls to lie along specific crystallographic directions, and each domain wall is assumed to have a $\pi/2$ bend. Integral-flux vortices (intermediate size orange dots) penetrate the superconductor along the straight sections of the domain walls.  At each bend, a bend flux penetrates the superconductor (large and small orange dots) and is fixed, via Eq.~(\ref{eqn:Bend_Angle_Flux}), to be a half-integer multiple of the flux quantum. All regions of localized flux are shown as if they had the same sign, as would be energetically favorable in the presence of a magnetic field applied along the $z$ axis. (color online)
\label{fig:lines_with_bend}}
\end{figure}

\begin{figure}
\includegraphics[width= .3\textwidth]{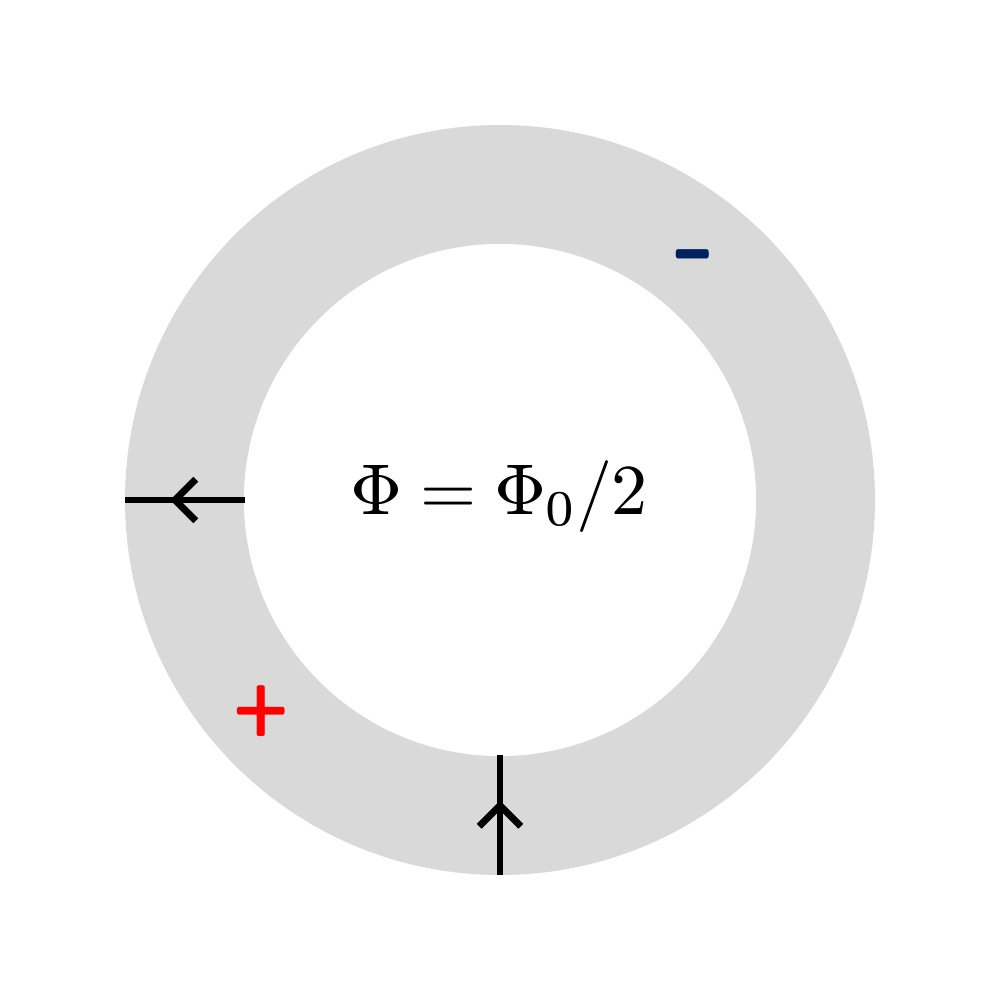}
\caption{Schematic depiction of an annular sample (shaded gray) the thickness and hieght of which is larger that the penetration depth. The annulus is crossed radially by a pair of domain walls oriented at $\pi/2$ relative to one another. Tetragonal $D_{4h}$ terms in the free energy are assumed to be large enough to pin the direction of the domain walls to crystallographically to lie along the specified directions. The minimum magnitude of the total flux through the hole would then be $\Phi_0/2$. (color online)
\label{fig:annulus}}
\end{figure}

We now describe three experimental scenarios in which it may prove possible to observe, via scanning magnetic microscopy, the phenomenon of sample penetration by nonintegral net magnetic flux associated with bends in walls separating regions of opposing chiral superconducting order.  These scenarios are depicted schematically in Figs.~\ref{fig:weak_pinning}, \ref{fig:lines_with_bend}, and~\ref{fig:annulus}. Augmenting the bend flux phenomenon discussed in the present work, it is known that domain walls are expected to produce Amp\`ere magnetic fields, resulting from currents that flow along the core of a domain walls; see Refs.~\cite{Volovik85, Matsumoto99}.  We emphasize that, even if there were a specific, microscopic reason for the magnitude of such currents to be reduced (cf., e.g., Refs.~\cite{Leggett06, Ashby09, Raghu10}), e.g., below currently detectable levels~\cite{Hicks10}, such a reduction would not affect the existence or magnitude of the bend fluxes discussed here. Thus, it is perhaps useful to regard bend fluxes as providing robust magnetic signatures of domain walls, and hence the form of superconductivity that spontaneously lacks time-reversal symmetry.

As a first example, consider a domain wall that intersects a physical surface of the superconducting system, the surface being oriented perpendicular to the $z$ axis. In this paper we are neglecting effects resulting from the finite height above the sample surface at which magnetic fields would typically be detected. (For a discussion of such effects see, e.g., Ref.~\cite{Bluhm07}.)\thinspace\  In addition, we envision domain walls to be pinned at generically located sites, e.g., by impurities.  In the limit in which the bulk terms in the free energy that break ${\rm SO}(2)_z$ symmetry are small (as can hold occur near $T_c$), the spatial arrangement of these pinning sites predominates in determining the bend angles that characterize a domain wall as it traverses the sample.
Assuming that these pinning sites are spaced further apart than the penetration depth, Eq.~(\ref{eqn:Bend_Angle_Flux}) indicates that these pinning locations would show up in scanning magnetometery as local regions of nonquantized flux penetrating the superconductor (see Fig.~\ref{fig:weak_pinning}).

We now outline a scenario specific to $\textrm{Sr}_2\textrm{RuO}_4$.  In both zero and nonzero in-plane magnetic fields, scanning magnetic imaging of $\textrm{Sr}_2\textrm{RuO}_4$ shows that vortices arrange themselves in line-like structures~\cite{Bjornsson05, Dolocan05, Dolocan06, Hicks10}.  One of the possible scenarios put forth to explain these structures is that the line-like structures are due to the binding of vortices to a parallel array of chiral domain walls~\cite{Sigrist99}.  However, to date, the line-like structures have not exhibited characteristics that would uniquely identify them as domain walls because, to within experimental uncertainty, the vortices (i.e., the local regions of penetrating magnetic field) were observed to have total fluxes that were integer multiples of $\Phi_0$, and Amp\`ere magnetic fields along the line-like features were not observed. The phenomenon of bend flux provides an additional route for determining whether the observed line-like structures are indeed associated with domain walls.  If it proves possible to prepare a sample (e.g., via a field-sweep procedure) so that the line-like features are bent then, if the line-like structures do indeed correspond to domain walls, bends would be accompanied by a nonintegral flux penetrating the superconductor (see Fig.~\ref{fig:lines_with_bend}).
The observation of nonintegral bend flux at a $\pi/2$ bend would provide further confirmation of p-wave pairing in $\textrm{Sr}_2\textrm{RuO}_4$ because, as noted in Sec.~\ref{sec:other-symms}, d-wave pairing would produce integer bend flux. However, the fact that Refs.~\cite{Bjornsson05,Dolocan05,Dolocan06} do not report regions of nonintegral
localized flux suggests that, in these experiments, if there are domain walls then they are aligned in parallel arrays, and thus are not bent.

A further consequence of domain walls should be evident in annular rings of broken time-reversal symmetry superconductors.  As the analysis leading to Eq.~(\ref{eqn:Bend_Angle_Flux}) is local only to the contour $\cal{C}$, and does not require inspection of the superconductivity near the domain-wall bend itself, it can be generalized to the case in which the bend is replaced by a hole in the superconductivity (see Fig.~\ref{fig:annulus}).  Recently, evidence for half-integer fluxoid behavior has been obtained in experiments on mesoscopic rings of $\textrm{Sr}_2\textrm{RuO}_4$ using cantilever torque magnetometry~\cite{Jang10}. However, in those experiments the half-integer fluxoid behavior was found to be accompanied by a small, rotationally invariant, in-plane component of the magnetization, and we are not aware of any reason why such a magnetization component would arise in the context of domain walls.

\section{Concluding remarks}
\label{sec:conclusion}

We have analyzed the properties of unconventional superconductors in which the superconducting state spontaneously breaks time-reversal symmetry and thus have the potential to exhibit domain walls that separate regions of opposing order-parameter chirality.
By employing an extension of the well-known London limit of the superconducting state, we have formulated an effective theory in terms of the topological variables that describe vortices and domain walls of the order parameter.
We have used this effective-theory formulation to show that localized near any bend in a domain wall through an angle $\Theta$, there is an associated net magnetic flux $\left((\Theta/\pi) + n\right)\Phi_0 $ (for some integer $n$)---provided the system can be taken to be rotationally invariant, crystallographically, about the $z$ axis.
We have also shown that this result for the flux near a domain-wall bend holds more generally.
Neither the London limit nor the regime of validity of the Ginzburg-Landau theory are required.  Rather it is sufficient for the following condition to hold: within regions of maximal chirality the two transformations, ${\rm SO}(2)_z$ rotations and ${\rm U}(1)$ gauge transformations of the superconduction order parameter, are degenerate transformations, in the sense that they have equivalent impacts on the state of the superconductivity.

We have addressed the issue of the relaxation of the assumption of crystallographic rotational invariance,
and its replacement by discrete rotational invariance.  In this regime we have found that the result for the bend flux continues to hold, but only for specific values of the bend angle, that are determined by the crystalline symmetry.  We have also sketched three candidate settings in which the interplay between chiral-domain-wall geometry and magnetic flux discussed in this paper might be observable, e.g., in experiments using scanned probe magnetic imaging.
The analysis that we have presented may be of use in determining the existence and distribution of domain walls in various superconducting materials such as $\rm{Sr}_2\rm{RuO}_4$, and may thus be of use in resolving the question of whether superconductivity in $\rm{Sr}_2\rm{RuO}_4$ does indeed spontaneously break time-reversal symmetry.
Aside from its intrinsic interest, resolving this question would, {\it inter alia\/}, be valuable in assessing the utility of materials such as these for exhibiting nonabelian phases and Majorana modes and, hence, the robustness with respect to decoherence that could prove useful for quantum information processing purposes.

\begin{acknowledgments}
We thank
Raffi Budakian, Suk-Bum Chung, Joonho Jang, Tony Leggett,
Catherine Kallin, Kathryn Moler, Michael Stone, Victor Vakaryuk,
and Shizhong Zhang for informative discussions.
This material is based upon work supported by the U.S.~Department of Energy, Division of Materials Sciences under Award No.~DE-FG02-07ER46453, through the Frederick Seitz Materials Research Laboratory at the University of Illinois at Urbana-Champaign.
\end{acknowledgments}

\appendix

\section{Free energy of a translationally invariant domain wall}
\label{sec:straight_dw}
In this appendix we use a variational approach to derive an estimate for the free energy per unit length $E_{\rm dw}$ of a translationally invariant domain wall, starting from the free energy $F_{\rm EL}$ (in the extended London limit), given in Eq.~(\ref{eqn:Feff}).  (Similar calculations can be found throughout the literature; see, e.g., Refs.~\cite{Volovik85,Sigrist89,Sigrist91,Sigrist99,Logoboy09,Ashby09,Bouhon10}.)\thinspace\
As we discussed in Sec.~\ref{sec:effective_theory}, $F_{\rm EL}$ contains two contributions:
one, $\fdwcore$, due to the core energy of a domain wall, which we estimate variationally;
and the other, the London term, that describes both the kinetic energy of supercurrents and the magnetic field energy.
In order to express $\fdwcore$ compactly, we define $(\alpha_1,\alpha_2)=(\gamma,\beta)$, and thus $\fdwcore$ become
\begin{subequations}
\begin{eqnarray}
    \fdwcore
    &=&
    \frac{1}{2} \nabla_a\,\alpha_i \Upsilon_{a i b j}\,\nabla_b \alpha_j + \frac{1}{8\Ldw^2}\cos^2\beta,
    \\
    \Upsilon_{a i c j}
    &:=&
    \frac{1}{4}
    \matI_{ab}
    \left(
      \begin{array}{cc}
        4 \cos^2 \beta & 0 \\
        0 & 1-\mu^2 \cos^2 \beta -\frac{\tau^2}{4} \\
      \end{array}
    \right)_{ij}
    \nonumber
    \\
    &&\qquad\qquad-
    \frac{\tau\cos\beta}{8}
    {\cal M}^{\gamma-(\pi/4)}_{ab}
    \left(
      \begin{array}{cc}
        4 & 0 \\
        0 & 1-2 \mu \\
      \end{array}
    \right)_{i j}
    \nonumber\\
    &&\qquad\qquad+
    \frac{\mu\sin 2\beta}{4}
    \matE_{ab}\,
    \matE_{ij},
\end{eqnarray}
\end{subequations}
where repeated indices $\{a,b,i,j\}$ are summed from 1 to 2.  The approach taken in this appendix is to evaluate independently the two contributions to Eq.~(\ref{eqn:Feff}), expressing separately the variational estimate for the core energy per unit length $E_{\rm core}$ and the London energy per unit length $E_{\rm L}$, and then to add these contributions to determine $E_{\rm dw}$.

To derive the variational estimate, we make the following assumptions for the spatial dependence of the $\gamma$ and $\beta$ fields transverse to the domain wall:
we take $\gamma$ to be constant and equal to $\dgfGamma$,
and we take $\beta(x)$ to be equal to $\beta_\ell(x):=2\tan^{-1}\tanh(x/2\ell)$,
where $\ell$ is a variational parameter specifying the width of the domain wall.
[To motivate the form $\beta_{\ell}$, we note that for $(\mu,\tau)=(0,0)$ and $\gamma$ constant, the term $\fdwcore$ reduces to
$\frac{1}{8}\vert{\bm\nabla}\beta\vert^{2}+\frac{\Ldw^{-2}}{8}\cos^{2}\beta$, and this form has the property of being stationary at $\beta_\ell(x)$, provided $\ell = \Ldw$.]

By using the variational assumptions for $\gamma$ and $\beta$ we obtain the following expressions for $E_{\rm core}$, which depends upon $\ell$ and $\dgfGamma$ as well as the angle $\phi$ [which specifies the direction $\hat{{\bm n}}=(\cos\phi,\sin\phi)$ normal to the domain wall]:
\begin{eqnarray}
    &&E_{\rm core}(\dgfGamma,\phi,\ell)
    \nonumber\\
    &&\qquad
    =\frac{1}{\ell}\bigg(\frac{\ell^2}{4 \Ldw^2}+\frac{1}{4}-\frac{\mu^2}{6}-\frac{\tau^2}{16}
    \nonumber\\
    &&\qquad\qquad
    +\left(\frac{\pi\,\mu \tau}{16} -\frac{\pi\,\tau}{32}\right)\cos(2 (\dgfGamma-\phi))\bigg).
\end{eqnarray}
By minimizing $E_{\rm core}$ with respect to $\ell$, we see that, in the extended London limit (for which $\Ldw$ tends to zero), the value of $\ell$ that makes $E_{\rm dw}$ stationary, would also tend to zero, provided the other energetic contribution, $E_{\rm L}$, does not force the stationary value of $\ell$ away from this result.  To see that indeed $E_{\rm L}$ does not do this, using the same varriational assumptions for $\gamma$ and $\beta$ we examine $E_{\rm L}$ expressed as power series in $\ell$ to ${\cal O}(\ell^{0})$:
\begin{eqnarray}
    &&E_{\rm L}(\dgfGamma,\phi,\ell)=
    \nonumber\\
    &&\quad\quad
    \frac{1}{\ell}
    \bigg(
    \frac{\mu^2}{6} +\frac{\tau^2}{32}
    - \frac{\pi\,\mu \tau}{16}\cos 2(\dgfGamma-\phi)
    \nonumber\\
    &&\quad\quad\quad\quad
    +\frac{ \tau^2}{32}\cos 4 (\dgfGamma -\phi)
    + {\cal O}(\ell)\bigg).
\end{eqnarray}
Combining the two terms, $E_{\rm core}$ and $E_{\rm L}$, we arrive at the following variational expression for the free energy per unit length of a translationally invariant domain wall:
\begin{eqnarray}
\label{eqn:DWline_energy}
    &&E_{{\rm dw}}(\dgfGamma,\phi,\ell)=
    \nonumber\\
    &&\quad\quad
    \frac{1}{4 \ell} \bigg( \frac{\ell^2}{\Ldw^2} + 1 -\frac{\tau^2}{8}
    -\frac{\pi\,\tau}{8}\cos 2(\dgfGamma-\phi)
    \nonumber\\
    &&\quad\quad\quad\quad
    +\frac{\tau^2}{8}\cos 4 (\dgfGamma -\phi)
    +{\cal O}(\ell)\bigg).
\end{eqnarray}
By minimizing $E_{\rm dw}$ with respect to $\ell$, and recalling that in the extended London limit $\Ldw$ is small, we find the stationary value of $\ell$ to be proportional to $\Ldw$, consistent with the assumption, just made, that $\ell$ is also small in the extended London limit.  Then, by replacing $\ell$ by its stationary value one obtains a value for $E_{{\rm dw}}$ having the following properties, some of which we make use in Sec.~\ref{sec:bending_domain_wall}:
(i)~it depends on $\dgfGamma$ and $\phi$ only through the combination $\dgfGamma-\phi$ and is $\pi$ periodic in this quantity;
(ii)~it is independent of $\mu$ (to leading order in $\Ldw$); and
(iii)~when $\tau \leq \pi/4$ the values of $\dgfGamma$ that minimizes $\min_{\ell} E_{\rm dw}$ are $\phi+n\pi$ (for integer $n$).

\begin{widetext}

\section{Free energy in terms of topological variables for the case of conventional superconductivity}
\label{sec:apdx_conventional_eff}

To motivate the derivation of the effective free energy for the topological variables given in Sec.~\ref{sec:effective_theory} of the main text, resulting in Eq.~(\ref{eqn:Feff}), we review in detail how it would proceed in the simpler setting of conventional superconductivity, and without employing the dual approach.  For a conventional superconductor, the order parameter is the complex scalar field $\psi(\posvec)$.  We assume that, in the absence of a magnetic field, the system is translationally and rotationally invariant, and we consider magnetic fields that are oriented along the $z$ direction and states of the superconductivity that are homogeneous in the $z$ direction.  In addition, we work with dependent and independent variables that have been rendered dimensionless via the rescalings given in Sec.~\ref{sec:PhenomUSC}.

With these assumptions we begin this derivation with the Ginzburg-Landau free energy per unit length of sample
\begin{equation}
F[\psi,{\bm A},{\bm H}]:=
    \frac{1}{2}\int d^2r \left(
    |({\bm \nabla}- i {\bm A}) \psi)|^2 +
    \frac{\kappa^2}{2} (|\psi|^2-1)^2 +
    |({\bm \nabla}\times {\bm A})- {\bm H}|^2 \right).
\end{equation}
In the London limit, in which $\kappa\to\infty$, the potential terms of this free energy fix the magnitude of $\psi$ to be unity.  Then, $\psi$ can be parametrized via a ${\rm U}(1)$ phase field $\theta(\posvec)$, so that $\psi(\posvec)\rightarrow\exp{i\theta(\posvec)}$.  Making this replacement in the free energy, we obtain the London form of the free energy, i.e.,
\begin{equation}
\label{eqn:convL}
    \frac{1}{2}\int d^2r \left(
    |{\bm \nabla} \theta - {\bm A}|^2 +
    |({\bm \nabla} \times {\bm A}) - {\bm H}|^2 \right).
\end{equation}
The two terms in the London free energy can be regarded as frustrating one another, energetically, as they impose competing demands on the ${\bm A}$ field.  The first term favors the transverse (i.e., divergence-free) part of ${\bm A}$ to be zero, the $\theta$ field can compensating for any longitudinal (i.e., curl-free) part; in contrast, the second term favors the transverse part of ${\bm A}$ to be nonzero.

For Type~II superconductors at magnetic fields above the lower critical field, a partial resolution to this frustration comes from the introduction of vortices, which alter the structure of the $\theta$ field: $\theta$ becomes multi-valued, and is singular in the cores of the vortices.  In particular, the expression ${\bm \nabla} \theta$ is not curl free and, correspondingly, has a transverse part.

To derive the effective free energy in terms of the appropriate topological variables (in this case, the density of vortices) one now decomposes the $\theta$ field into a smooth, single-valued part $\theta_{\rm sm}$ and a part $\theta_{\rm v}$ that contains the vortex singularities, so that $\theta = \theta_{\rm sm} + \theta_{\rm v}$. Next, one seeks to eliminate $\theta_{\rm sm}$ from the free energy by setting it to the value that makes the
free energy stationary.  As the only term in the free energy that depends on $\theta_{\rm sm}$ is the one corresponding to the kinetic energy of the supercurrents [i.e., the former term in Eq.~(\ref{eqn:convL})], for the issue of stationarity one need only consider this term.  Expanding it and integrating by parts, gives
\begin{equation}
\label{eqn:thetasm}
    \int d^2r \bigg(
    \frac{1}{2}\theta_{\rm sm}(-\nabla^2)\theta_{\rm sm}
    - \theta_{\rm sm} {\bm \nabla}
    \cdot ({\bm \nabla} \theta_{\rm v} - {\bm A})
    +\frac{1}{2}|{\bm \nabla} \theta_{\rm v} - {\bm A}|^2 \bigg).
\end{equation}
Then, using the Green function for the Laplacian in two dimensions, which obeys
$-\nabla^2 G(\posvec)= \delta(\posvec)$ and reads
$G(\posvec) = -({2 \pi})^{-1}\ln|\posvec|$,
one finds that at stationarity $\theta_{\rm sm}$ is given by
\begin{equation}
    \bar{\theta}_{\rm sm}(\posvec^{\prime}) =
    -\int d^{2}r\,G(\posvec^{\prime}-\posvec){\bm\nabla}\cdot
    \big({\bm\nabla}\theta_{\rm v}(\posvec) - {\bm A}(\posvec)\big).
\end{equation}
By inserting $\bar{\theta}_{\rm sm}$ into Eq.~(\ref{eqn:thetasm}) and using the defining equation for $G(\posvec)$ to express the last term of Eq.~(\ref{eqn:thetasm}) in terms of $G$, one obtains for the kinetic energy of the supercurrent
\def\localbig{\Big}
\begin{eqnarray}
    \frac{1}{2}\int d^2r\,d^2r'\,
    \bigg(%
    &-&\localbig(\nabla_a(\nabla_a \theta_{\rm v} - A_a)({\bm r})\localbig)\,
    G({\bm r} - {\bm r}')
    \localbig(\nabla'_b(\nabla'_b \theta_{\rm v} - A_a)({\bm r}')\localbig)
    \nonumber\\
    &+&\localbig((\nabla_a \theta_{\rm v} - A_a)({\bm r})\localbig)
    \localbig(-\nabla^2 G({\bm r} - {\bm r}')\localbig)
    \localbig((\nabla'_a \theta_{\rm v} - A_a)({\bm r}')\localbig)
    \bigg).
\end{eqnarray}
The two terms in this equation have similar structure, and integration by parts allows them to be expressed as
\begin{eqnarray}
    &&\frac{1}{2}\int d^2r\,d^2r'\,
    \localbig(I_{ab}I_{cd} - I_{ad} I_{bc}\localbig)
    \localbig(\nabla_a \theta_{\rm v} - A_a\localbig)({\bm r})\,
    \localbig(\nabla_b \nabla_c G({\bm r} - {\bm r}')\localbig)\,
    \localbig(\nabla'_d \theta_{\rm v} - A_d)({\bm r}')\localbig).
\end{eqnarray}
Next, by using the elementary tensor identity
\begin{equation}
I_{ab}\,I_{cd} - I_{ad}\,I_{bc} = E_{ac}\,E_{bd}
\end{equation}
and integrating by parts, the suprcurrent kinetic energy becomes
\begin{eqnarray}
    \label{eq:transsuperdensity}
    &&\frac{1}{2}\int d^2r\,d^2r'\,
    \bigg(
    E_{ab}\,
    \localbig(\nabla_a(\nabla_b \theta_{\rm v} - A_a)({\bm r})\localbig)\,
    G({\bm r} - {\bm r}')\,
    E_{cd}\,
    \localbig(\nabla'_c(\nabla'_d \theta_{\rm v} - A_a)({\bm r}')\localbig)
    \bigg).
\end{eqnarray}
This form shows that the elimination of the smooth part of $\theta$ creates a long-ranged interaction for the curl of ${\bm \nabla} \theta_{\rm v} - {\bm A}$.  This free energy can readily be shown to be equivalent to Eq.~(\ref{eqn:TransverseCurrents}), and thus to describe the kinetic energy of the transverse part of the supercurrent.  Equation~(\ref{eq:transsuperdensity}) features the curl of the gradient of the multi-valued function $\theta_{\rm v}$, which is a combination that isolates the $\delta$-function contributions from the singularities in the vortex cores, so that
\begin{equation}
    E_{ab}\nabla_a \nabla_b \theta_{\rm v} = 2\pi\,\rho_{\rm v},
\end{equation}
where $\rho_{\rm v}(\posvec):=\sum q_\nu \delta(\posvec-{\bm R}_{\nu})$ defines the vortex density in terms of the vortex locations $\{{\bm R}_{\nu}\}$ and vorticity $\{ q_\nu \}$.  In particular, one sees that owing to the vortices the gradient of $\theta$ can posses a transverse part, and this can partially relieve the frustration of ${\bm A}$ inherent in the London free energy.

To proceed further with the derivation of the effective free energy in terms of vortex variables, one now considers the full London free energy, Eq.~(\ref{eqn:convL}), which, in terms of the total magnetic field $B=E_{ab}\nabla_a A_b$, reads
\begin{eqnarray}
\label{eqn:Bnonlocal}
    &&\frac{1}{2}\int d^2r\,d^2r'\,
    (2 \pi \rho_{\rm v} - B)({\bm r})\,
    G({\bm r} - {\bm r}')\,
    (2 \pi \rho_{\rm v} - B)({\bm r}')
    +\frac{1}{2}\int d^2r\,(B-H)^2.
\end{eqnarray}
Note that we have omitted a constant contribution resulting from the suppression of the magnitude of the order parameter within the core of each vortex, as it is negligibly small, relative to the kinetic and field energies,  in the London limit.

The next step is to eliminate the magnetic field from the free energy by setting it to its stationary value $\bar{B}$ which, from Eq.~(\ref{eqn:Bnonlocal}), one sees is
\begin{equation}
\label{eqn:nonlocalsolution}
\bar{B}(\posvec)=
    \int d^2r'\,
    \Big(G+\delta\Big)^{-1}(\posvec-\posvec')\,
\left(
    H(\posvec')+
    2\pi\int d^2r''\,
    G(\posvec'-\posvec'')\,\rho(\posvec'')
\right),
\end{equation}
where $\big(G+\delta\big)^{-1}(\posvec-\posvec')$ is the inverse of the kernel
$G(\posvec-\posvec')+\delta(\posvec-\posvec')$.  It is convenient to adopt a schematic notation in which one suppresses integral signs and dependences on spatial variables, in which case the result for $\bar{B}$ reads
\begin{equation}
    \bar{B} = (G+\delta)^{-1}
    (H+2\pi\,G\,\rho).
\end{equation}
Replacing $B$ by $\bar{B}$ in Eq.~(\ref{eqn:Bnonlocal}) then yields the following expression for the free energy:
\begin{eqnarray}
\label{eqn:Bdirect}
    \frac{1}{2} \left(
    (2\pi \rho)\left(G - G (G+\delta)^{-1} G\right)(2 \pi \rho) +
     H\left(\delta - (G+\delta)^{-1}\right) H
    -H(G+\delta)^{-1} G (2 \pi \rho)
    -(2 \pi \rho) G (G+\delta)^{-1} H\right).
\end{eqnarray}
It is straightforward to see that each of the four integral kernels in this formula is the Green function for the Helmholtz operator in two dimensions, which obeys $(-\nabla^2 +1){\cal G}(\posvec) = \delta(\posvec)$, and is given by ${\cal G}(\posvec) = ({2 \pi})^{-1}K_0(|\posvec|)$, where $K_0(x)$ is a modified Bessel function.  To exemplify this one can apply the following elementary manipulations to the kernel of the first term:
\begin{eqnarray}
    G-G(G+\delta)^{-1}\,G =
    G\,\Big(\delta - \big(-\nabla^2 (G+\delta)\big)^{-1}\Big) =
    G\,\Big((-\nabla^2 + \delta){\cal G} - {\cal G}\Big) =
    G\,\big(-\nabla^2\big)\,{\cal G} = {\cal G}.
\end{eqnarray}
By similarly simplifying the remaining kernels in Eq.~(\ref{eqn:Bdirect}) one completes the derivation of the effective free energy in terms of the vortex density $\rho_{\rm v}$ and the applied field $H$, arriving at the result
\begin{equation}
    \frac{1}{4\pi}\int d^2r\,d^2r'\,
    \localbig(2 \pi \rho_{\rm v} - H\localbig)({\bm r})\,
    K_0(|{\bm r} - {\bm r}'|)\,
    \localbig(2 \pi \rho_{\rm v} - H\localbig)({\bm r}'),
\end{equation}
which is the analog for conventional superconductivity of the unconventional superconductivity formula Eq.~(\ref{eqn:Feff}).

\end{widetext}

\bibliography{WallBend}

\begin{thebibliography}{65}%
\makeatletter
\providecommand \@ifxundefined [1]{%
 \@ifx{#1\undefined}
}%
\providecommand \@ifnum [1]{%
 \ifnum #1\expandafter \@firstoftwo
 \else \expandafter \@secondoftwo
 \fi
}%
\providecommand \@ifx [1]{%
 \ifx #1\expandafter \@firstoftwo
 \else \expandafter \@secondoftwo
 \fi
}%
\providecommand \natexlab [1]{#1}%
\providecommand \enquote  [1]{``#1''}%
\providecommand \bibnamefont  [1]{#1}%
\providecommand \bibfnamefont [1]{#1}%
\providecommand \citenamefont [1]{#1}%
\providecommand \href@noop [0]{\@secondoftwo}%
\providecommand \href [0]{\begingroup \@sanitize@url \@href}%
\providecommand \@href[1]{\@@startlink{#1}\@@href}%
\providecommand \@@href[1]{\endgroup#1\@@endlink}%
\providecommand \@sanitize@url [0]{\catcode `\\12\catcode `\$12\catcode
  `\&12\catcode `\#12\catcode `\^12\catcode `\_12\catcode `\%12\relax}%
\providecommand \@@startlink[1]{}%
\providecommand \@@endlink[0]{}%
\providecommand \url  [0]{\begingroup\@sanitize@url \@url }%
\providecommand \@url [1]{\endgroup\@href {#1}{\urlprefix }}%
\providecommand \urlprefix  [0]{URL }%
\providecommand \Eprint [0]{\href }%
\providecommand \doibase [0]{http://dx.doi.org/}%
\providecommand \selectlanguage [0]{\@gobble}%
\providecommand \bibinfo  [0]{\@secondoftwo}%
\providecommand \bibfield  [0]{\@secondoftwo}%
\providecommand \translation [1]{[#1]}%
\providecommand \BibitemOpen [0]{}%
\providecommand \bibitemStop [0]{}%
\providecommand \bibitemNoStop [0]{.\EOS\space}%
\providecommand \EOS [0]{\spacefactor3000\relax}%
\providecommand \BibitemShut  [1]{\csname bibitem#1\endcsname}%
\let\auto@bib@innerbib\@empty
\bibitem [{\citenamefont {Mackenzie}\ and\ \citenamefont
  {Maeno}(2003)}]{Mackenzie03}%
  \BibitemOpen
  \bibfield  {author} {\bibinfo {author} {\bibfnamefont {A.~P.}\ \bibnamefont
  {Mackenzie}}\ and\ \bibinfo {author} {\bibfnamefont {Y.}~\bibnamefont
  {Maeno}},\ }\href@noop {} {\bibfield  {journal} {\bibinfo  {journal} {Rev.
  Mod. Phys.}\ }\textbf {\bibinfo {volume} {75}},\ \bibinfo {pages} {657}
  (\bibinfo {year} {2003})}\BibitemShut {NoStop}%
\bibitem [{\citenamefont {Rice}\ and\ \citenamefont {Sigrist}(1995)}]{Rice95}%
  \BibitemOpen
  \bibfield  {author} {\bibinfo {author} {\bibfnamefont {T.~M.}\ \bibnamefont
  {Rice}}\ and\ \bibinfo {author} {\bibfnamefont {M.}~\bibnamefont {Sigrist}},\
  }\href@noop {} {\bibfield  {journal} {\bibinfo  {journal} {J. Phys. Cond.
  Mat.}\ ,\ \bibinfo {pages} {L643}} (\bibinfo {year} {1995})}\BibitemShut
  {NoStop}%
\bibitem [{\citenamefont {Ishida}\ \emph {et~al.}(1998)\citenamefont {Ishida},
  \citenamefont {Mukuda}, \citenamefont {Kitaoka}, \citenamefont {Asayama},
  \citenamefont {Mao}, \citenamefont {Mori},\ and\ \citenamefont
  {Maeno}}]{Ishida98}%
  \BibitemOpen
  \bibfield  {author} {\bibinfo {author} {\bibfnamefont {K.}~\bibnamefont
  {Ishida}}, \bibinfo {author} {\bibfnamefont {H.}~\bibnamefont {Mukuda}},
  \bibinfo {author} {\bibfnamefont {Y.}~\bibnamefont {Kitaoka}}, \bibinfo
  {author} {\bibfnamefont {K.}~\bibnamefont {Asayama}}, \bibinfo {author}
  {\bibfnamefont {Z.~Q.}\ \bibnamefont {Mao}}, \bibinfo {author} {\bibfnamefont
  {Y.}~\bibnamefont {Mori}}, \ and\ \bibinfo {author} {\bibfnamefont
  {Y.}~\bibnamefont {Maeno}},\ }\href@noop {} {\bibfield  {journal} {\bibinfo
  {journal} {Nature}\ }\textbf {\bibinfo {volume} {396}},\ \bibinfo {pages}
  {658} (\bibinfo {year} {1998})}\BibitemShut {NoStop}%
\bibitem [{\citenamefont {Nelson}\ \emph {et~al.}(2004)\citenamefont {Nelson},
  \citenamefont {Mao}, \citenamefont {Maeno},\ and\ \citenamefont
  {Liu}}]{Nelson04}%
  \BibitemOpen
  \bibfield  {author} {\bibinfo {author} {\bibfnamefont {K.~D.}\ \bibnamefont
  {Nelson}}, \bibinfo {author} {\bibfnamefont {Z.~Q.}\ \bibnamefont {Mao}},
  \bibinfo {author} {\bibfnamefont {Y.}~\bibnamefont {Maeno}}, \ and\ \bibinfo
  {author} {\bibfnamefont {Y.}~\bibnamefont {Liu}},\ }\href@noop {} {\bibfield
  {journal} {\bibinfo  {journal} {Science}\ }\textbf {\bibinfo {volume}
  {306}},\ \bibinfo {pages} {1151} (\bibinfo {year} {2004})}\BibitemShut
  {NoStop}%
\bibitem [{\citenamefont {Jang}\ \emph {et~al.}()\citenamefont {Jang},
  \citenamefont {Ferguson}, \citenamefont {Vakaryuk}, \citenamefont {Budakian},
  \citenamefont {Chung}, \citenamefont {Goldbart},\ and\ \citenamefont
  {Maeno}}]{Jang10}%
  \BibitemOpen
  \bibfield  {author} {\bibinfo {author} {\bibfnamefont {J.}~\bibnamefont
  {Jang}}, \bibinfo {author} {\bibfnamefont {D.}~\bibnamefont {Ferguson}},
  \bibinfo {author} {\bibfnamefont {V.}~\bibnamefont {Vakaryuk}}, \bibinfo
  {author} {\bibfnamefont {R.}~\bibnamefont {Budakian}}, \bibinfo {author}
  {\bibfnamefont {S.}~\bibnamefont {Chung}}, \bibinfo {author} {\bibfnamefont
  {P.}~\bibnamefont {Goldbart}}, \ and\ \bibinfo {author} {\bibfnamefont
  {Y.}~\bibnamefont {Maeno}},\ }\href@noop {} {\bibinfo  {journal} {(2010,
  unpublished)}\ }\BibitemShut {NoStop}%
\bibitem [{\citenamefont {Kopnin}\ and\ \citenamefont
  {Salomaa}(1991)}]{Kopnin91}%
  \BibitemOpen
\bibfield  {journal} {  }\bibfield  {author} {\bibinfo {author} {\bibfnamefont
  {N.~B.}\ \bibnamefont {Kopnin}}\ and\ \bibinfo {author} {\bibfnamefont
  {M.~M.}\ \bibnamefont {Salomaa}},\ }\href@noop {} {\bibfield  {journal}
  {\bibinfo  {journal} {Phys. Rev. B}\ }\textbf {\bibinfo {volume} {44}},\
  \bibinfo {pages} {9667} (\bibinfo {year} {1991})}\BibitemShut {NoStop}%
\bibitem [{\citenamefont {Read}\ and\ \citenamefont {Green}(2000)}]{Read00}%
  \BibitemOpen
  \bibfield  {author} {\bibinfo {author} {\bibfnamefont {N.}~\bibnamefont
  {Read}}\ and\ \bibinfo {author} {\bibfnamefont {D.}~\bibnamefont {Green}},\
  }\href@noop {} {\bibfield  {journal} {\bibinfo  {journal} {Phys. Rev. B}\
  }\textbf {\bibinfo {volume} {61}},\ \bibinfo {pages} {10267} (\bibinfo {year}
  {2000})}\BibitemShut {NoStop}%
\bibitem [{\citenamefont {Kitaev}(2003)}]{Kitaev03}%
  \BibitemOpen
  \bibfield  {author} {\bibinfo {author} {\bibfnamefont {A.~Y.}\ \bibnamefont
  {Kitaev}},\ }\href@noop {} {\bibfield  {journal} {\bibinfo  {journal} {Ann.
  Phys. (NY)}\ }\textbf {\bibinfo {volume} {303}},\ \bibinfo {pages} {2}
  (\bibinfo {year} {2003})}\BibitemShut {NoStop}%
\bibitem [{\citenamefont {Nayak}\ \emph {et~al.}(2008)\citenamefont {Nayak},
  \citenamefont {Simon}, \citenamefont {Stern}, \citenamefont {Freedman},\ and\
  \citenamefont {Sarma}}]{Nayak08}%
  \BibitemOpen
  \bibfield  {author} {\bibinfo {author} {\bibfnamefont {C.}~\bibnamefont
  {Nayak}}, \bibinfo {author} {\bibfnamefont {S.~H.}\ \bibnamefont {Simon}},
  \bibinfo {author} {\bibfnamefont {A.}~\bibnamefont {Stern}}, \bibinfo
  {author} {\bibfnamefont {M.}~\bibnamefont {Freedman}}, \ and\ \bibinfo
  {author} {\bibfnamefont {S.~D.}\ \bibnamefont {Sarma}},\ }\href@noop {}
  {\bibfield  {journal} {\bibinfo  {journal} {Rev. Mod. Phys.}\ }\textbf
  {\bibinfo {volume} {80}},\ \bibinfo {pages} {1083} (\bibinfo {year}
  {2008})}\BibitemShut {NoStop}%
\bibitem [{\citenamefont {Yoshioka}\ and\ \citenamefont
  {Miyake}(2009)}]{Yoshioka09}%
  \BibitemOpen
  \bibfield  {author} {\bibinfo {author} {\bibfnamefont {Y.}~\bibnamefont
  {Yoshioka}}\ and\ \bibinfo {author} {\bibfnamefont {K.}~\bibnamefont
  {Miyake}},\ }\href@noop {} {\bibfield  {journal} {\bibinfo  {journal} {J.
  Phys. Soc. Jpn.}\ }\textbf {\bibinfo {volume} {78}},\ \bibinfo {pages}
  {074701} (\bibinfo {year} {2009})}\BibitemShut {NoStop}%
\bibitem [{\citenamefont {Raghu}\ \emph {et~al.}(2010)\citenamefont {Raghu},
  \citenamefont {Kapitulnik},\ and\ \citenamefont {Kivelson}}]{Raghu10}%
  \BibitemOpen
  \bibfield  {author} {\bibinfo {author} {\bibfnamefont {S.}~\bibnamefont
  {Raghu}}, \bibinfo {author} {\bibfnamefont {A.}~\bibnamefont {Kapitulnik}}, \
  and\ \bibinfo {author} {\bibfnamefont {S.~A.}\ \bibnamefont {Kivelson}},\
  }\href@noop {} {\bibfield  {journal} {\bibinfo  {journal} {Phys. Rev. Lett.}\
  }\textbf {\bibinfo {volume} {105}},\ \bibinfo {pages} {136401} (\bibinfo
  {year} {2010})}\BibitemShut {NoStop}%
\bibitem [{\citenamefont {Luke}\ \emph {et~al.}(1998)\citenamefont {Luke},
  \citenamefont {Fudamoto}, \citenamefont {Kojima}, \citenamefont {Larkin},
  \citenamefont {Merrin}, \citenamefont {Nachumi}, \citenamefont {Uemura},
  \citenamefont {Maeno}, \citenamefont {Z.~Q.~Mao}, \citenamefont {Nakamura},\
  and\ \citenamefont {Sigrist}}]{Luke98}%
  \BibitemOpen
  \bibfield  {author} {\bibinfo {author} {\bibfnamefont {G.~M.}\ \bibnamefont
  {Luke}}, \bibinfo {author} {\bibfnamefont {Y.}~\bibnamefont {Fudamoto}},
  \bibinfo {author} {\bibfnamefont {K.~M.}\ \bibnamefont {Kojima}}, \bibinfo
  {author} {\bibfnamefont {M.~I.}\ \bibnamefont {Larkin}}, \bibinfo {author}
  {\bibfnamefont {J.}~\bibnamefont {Merrin}}, \bibinfo {author} {\bibfnamefont
  {B.}~\bibnamefont {Nachumi}}, \bibinfo {author} {\bibfnamefont {Y.~J.}\
  \bibnamefont {Uemura}}, \bibinfo {author} {\bibfnamefont {Y.}~\bibnamefont
  {Maeno}}, \bibinfo {author} {\bibfnamefont {Y.~M.}\ \bibnamefont
  {Z.~Q.~Mao}}, \bibinfo {author} {\bibfnamefont {H.}~\bibnamefont {Nakamura}},
  \ and\ \bibinfo {author} {\bibfnamefont {M.}~\bibnamefont {Sigrist}},\
  }\href@noop {} {\bibfield  {journal} {\bibinfo  {journal} {Nature}\ }\textbf
  {\bibinfo {volume} {394}},\ \bibinfo {pages} {558} (\bibinfo {year}
  {1998})}\BibitemShut {NoStop}%
\bibitem [{\citenamefont {Xia}\ \emph {et~al.}(2006)\citenamefont {Xia},
  \citenamefont {Maeno}, \citenamefont {Beyersdorf}, \citenamefont {Fejer},\
  and\ \citenamefont {Kapitulnik}}]{Xia06}%
  \BibitemOpen
  \bibfield  {author} {\bibinfo {author} {\bibfnamefont {J.}~\bibnamefont
  {Xia}}, \bibinfo {author} {\bibfnamefont {Y.}~\bibnamefont {Maeno}}, \bibinfo
  {author} {\bibfnamefont {P.~T.}\ \bibnamefont {Beyersdorf}}, \bibinfo
  {author} {\bibfnamefont {M.~M.}\ \bibnamefont {Fejer}}, \ and\ \bibinfo
  {author} {\bibfnamefont {A.}~\bibnamefont {Kapitulnik}},\ }\href@noop {}
  {\bibfield  {journal} {\bibinfo  {journal} {Phys. Rev. Lett.}\ }\textbf
  {\bibinfo {volume} {97}},\ \bibinfo {pages} {167002} (\bibinfo {year}
  {2006})}\BibitemShut {NoStop}%
\bibitem [{\citenamefont {Kallin}\ and\ \citenamefont
  {Berlinsky}(2009)}]{Kallin09}%
  \BibitemOpen
  \bibfield  {author} {\bibinfo {author} {\bibfnamefont {C.}~\bibnamefont
  {Kallin}}\ and\ \bibinfo {author} {\bibfnamefont {A.~J.}\ \bibnamefont
  {Berlinsky}},\ }\href@noop {} {\bibfield  {journal} {\bibinfo  {journal} {J.
  Phys. Cond. Mat.}\ }\textbf {\bibinfo {volume} {21}},\ \bibinfo {pages}
  {164210} (\bibinfo {year} {2009})}\BibitemShut {NoStop}%
\bibitem [{\citenamefont {Matsumoto}\ and\ \citenamefont
  {Sigrist}(1999)}]{Matsumoto99}%
  \BibitemOpen
  \bibfield  {author} {\bibinfo {author} {\bibfnamefont {M.}~\bibnamefont
  {Matsumoto}}\ and\ \bibinfo {author} {\bibfnamefont {M.}~\bibnamefont
  {Sigrist}},\ }\href@noop {} {\bibfield  {journal} {\bibinfo  {journal} {J.
  Phys. Soc. Jpn.}\ }\textbf {\bibinfo {volume} {68}},\ \bibinfo {pages} {994}
  (\bibinfo {year} {1999})}\BibitemShut {NoStop}%
\bibitem [{\citenamefont {Bj$\ddot{{\rm o}}$rnsson}\ \emph
  {et~al.}(2005)\citenamefont {Bj$\ddot{{\rm o}}$rnsson}, \citenamefont
  {Maeno}, \citenamefont {Huber},\ and\ \citenamefont {Moler}}]{Bjornsson05}%
  \BibitemOpen
  \bibfield  {author} {\bibinfo {author} {\bibfnamefont {P.~G.}\ \bibnamefont
  {Bj$\ddot{{\rm o}}$rnsson}}, \bibinfo {author} {\bibfnamefont
  {Y.}~\bibnamefont {Maeno}}, \bibinfo {author} {\bibfnamefont {M.~E.}\
  \bibnamefont {Huber}}, \ and\ \bibinfo {author} {\bibfnamefont {K.~A.}\
  \bibnamefont {Moler}},\ }\href@noop {} {\bibfield  {journal} {\bibinfo
  {journal} {Phys. Rev. B}\ }\textbf {\bibinfo {volume} {72}},\ \bibinfo
  {pages} {012504} (\bibinfo {year} {2005})}\BibitemShut {NoStop}%
\bibitem [{\citenamefont {Kirtley}\ \emph {et~al.}(2007)\citenamefont
  {Kirtley}, \citenamefont {Kallin}, \citenamefont {Hicks}, \citenamefont
  {Kim}, \citenamefont {Liu}, \citenamefont {Moler}, \citenamefont {Maeno},\
  and\ \citenamefont {Nelson}}]{Kirtley07}%
  \BibitemOpen
  \bibfield  {author} {\bibinfo {author} {\bibfnamefont {J.~R.}\ \bibnamefont
  {Kirtley}}, \bibinfo {author} {\bibfnamefont {C.}~\bibnamefont {Kallin}},
  \bibinfo {author} {\bibfnamefont {C.~W.}\ \bibnamefont {Hicks}}, \bibinfo
  {author} {\bibfnamefont {E.~A.}\ \bibnamefont {Kim}}, \bibinfo {author}
  {\bibfnamefont {Y.}~\bibnamefont {Liu}}, \bibinfo {author} {\bibfnamefont
  {K.~A.}\ \bibnamefont {Moler}}, \bibinfo {author} {\bibfnamefont
  {Y.}~\bibnamefont {Maeno}}, \ and\ \bibinfo {author} {\bibfnamefont {K.~D.}\
  \bibnamefont {Nelson}},\ }\href@noop {} {\bibfield  {journal} {\bibinfo
  {journal} {Phys. Rev. B}\ }\textbf {\bibinfo {volume} {76}},\ \bibinfo
  {pages} {014526} (\bibinfo {year} {2007})}\BibitemShut {NoStop}%
\bibitem [{\citenamefont {Hicks}\ \emph {et~al.}(2010)\citenamefont {Hicks},
  \citenamefont {Kirtley}, \citenamefont {Lippman}, \citenamefont {Koshnick},
  \citenamefont {Huber}, \citenamefont {Maeno}, \citenamefont {Yuhasz},
  \citenamefont {Maple},\ and\ \citenamefont {Moler}}]{Hicks10}%
  \BibitemOpen
  \bibfield  {author} {\bibinfo {author} {\bibfnamefont {C.~W.}\ \bibnamefont
  {Hicks}}, \bibinfo {author} {\bibfnamefont {J.~R.}\ \bibnamefont {Kirtley}},
  \bibinfo {author} {\bibfnamefont {T.~M.}\ \bibnamefont {Lippman}}, \bibinfo
  {author} {\bibfnamefont {N.~C.}\ \bibnamefont {Koshnick}}, \bibinfo {author}
  {\bibfnamefont {M.~E.}\ \bibnamefont {Huber}}, \bibinfo {author}
  {\bibfnamefont {Y.}~\bibnamefont {Maeno}}, \bibinfo {author} {\bibfnamefont
  {W.~M.}\ \bibnamefont {Yuhasz}}, \bibinfo {author} {\bibfnamefont {M.~B.}\
  \bibnamefont {Maple}}, \ and\ \bibinfo {author} {\bibfnamefont {K.~A.}\
  \bibnamefont {Moler}},\ }\href@noop {} {\bibfield  {journal} {\bibinfo
  {journal} {Phys. Rev. B}\ }\textbf {\bibinfo {volume} {81}},\ \bibinfo
  {pages} {214501} (\bibinfo {year} {2010})}\BibitemShut {NoStop}%
\bibitem [{\citenamefont {Volovik}\ and\ \citenamefont
  {Gor'kov}(1985)}]{Volovik85}%
  \BibitemOpen
  \bibfield  {author} {\bibinfo {author} {\bibfnamefont {G.}~\bibnamefont
  {Volovik}}\ and\ \bibinfo {author} {\bibfnamefont {L.}~\bibnamefont
  {Gor'kov}},\ }\href@noop {} {\bibfield  {journal} {\bibinfo  {journal}
  {JETP}\ }\textbf {\bibinfo {volume} {61}},\ \bibinfo {pages} {843} (\bibinfo
  {year} {1985})}\BibitemShut {NoStop}%
\bibitem [{\citenamefont {London}\ and\ \citenamefont
  {London}(1935)}]{London35}%
  \BibitemOpen
  \bibfield  {author} {\bibinfo {author} {\bibfnamefont {F.}~\bibnamefont
  {London}}\ and\ \bibinfo {author} {\bibfnamefont {H.}~\bibnamefont
  {London}},\ }\href@noop {} {\bibfield  {journal} {\bibinfo  {journal} {Proc.
  R. Soc. Lond. A}\ }\textbf {\bibinfo {volume} {149}},\ \bibinfo {pages} {71}
  (\bibinfo {year} {1935})}\BibitemShut {NoStop}%
\bibitem [{\citenamefont {Ginzburg}\ and\ \citenamefont
  {Landau}(1950)}]{Ginzburg50}%
  \BibitemOpen
  \bibfield  {author} {\bibinfo {author} {\bibfnamefont {V.}~\bibnamefont
  {Ginzburg}}\ and\ \bibinfo {author} {\bibfnamefont {L.}~\bibnamefont
  {Landau}},\ }\href@noop {} {\bibfield  {journal} {\bibinfo  {journal} {Zh.
  Eksp. Teor. Fiz.}\ }\textbf {\bibinfo {volume} {20}},\ \bibinfo {pages}
  {1064} (\bibinfo {year} {1950})}\BibitemShut {NoStop}%
\bibitem [{\citenamefont {Bardeen}\ \emph {et~al.}(1957)\citenamefont
  {Bardeen}, \citenamefont {Cooper},\ and\ \citenamefont
  {Schrieffer}}]{Bardeen57}%
  \BibitemOpen
  \bibfield  {author} {\bibinfo {author} {\bibfnamefont {J.}~\bibnamefont
  {Bardeen}}, \bibinfo {author} {\bibfnamefont {L.~N.}\ \bibnamefont {Cooper}},
  \ and\ \bibinfo {author} {\bibfnamefont {J.~R.}\ \bibnamefont {Schrieffer}},\
  }\href@noop {} {\bibfield  {journal} {\bibinfo  {journal} {Phys. Rev.}\
  }\textbf {\bibinfo {volume} {108}},\ \bibinfo {pages} {1175} (\bibinfo {year}
  {1957})}\BibitemShut {NoStop}%
\bibitem [{\citenamefont {Sigrist}\ and\ \citenamefont
  {Ueda}(1991)}]{Sigrist91}%
  \BibitemOpen
  \bibfield  {author} {\bibinfo {author} {\bibfnamefont {M.}~\bibnamefont
  {Sigrist}}\ and\ \bibinfo {author} {\bibfnamefont {K.}~\bibnamefont {Ueda}},\
  }\href@noop {} {\bibfield  {journal} {\bibinfo  {journal} {Rev. Mod. Phys.}\
  }\textbf {\bibinfo {volume} {63}},\ \bibinfo {pages} {239} (\bibinfo {year}
  {1991})}\BibitemShut {NoStop}%
\bibitem [{\citenamefont {Mineev}\ and\ \citenamefont
  {Samokhin}(1999)}]{Mineev99}%
  \BibitemOpen
  \bibfield  {author} {\bibinfo {author} {\bibfnamefont {V.~P.}\ \bibnamefont
  {Mineev}}\ and\ \bibinfo {author} {\bibfnamefont {K.~V.}\ \bibnamefont
  {Samokhin}},\ }\href@noop {} {\emph {\bibinfo {title} {Introduction to
  Unconventional Superconductivity}}}\ (\bibinfo  {publisher} {Gordon and
  Breach},\ \bibinfo {address} {Amsterdam},\ \bibinfo {year}
  {1999})\BibitemShut {NoStop}%
\bibitem [{\citenamefont {Sigrist}\ \emph {et~al.}(1989)\citenamefont
  {Sigrist}, \citenamefont {Rice},\ and\ \citenamefont {Ueda}}]{Sigrist89}%
  \BibitemOpen
  \bibfield  {author} {\bibinfo {author} {\bibfnamefont {M.}~\bibnamefont
  {Sigrist}}, \bibinfo {author} {\bibfnamefont {T.~M.}\ \bibnamefont {Rice}}, \
  and\ \bibinfo {author} {\bibfnamefont {K.}~\bibnamefont {Ueda}},\ }\href@noop
  {} {\bibfield  {journal} {\bibinfo  {journal} {Phys. Rev. Lett.}\ }\textbf
  {\bibinfo {volume} {63}},\ \bibinfo {pages} {1727} (\bibinfo {year}
  {1989})}\BibitemShut {NoStop}%
\bibitem [{\citenamefont {Sigrist}\ and\ \citenamefont
  {Agterberg}(1999)}]{Sigrist99}%
  \BibitemOpen
  \bibfield  {author} {\bibinfo {author} {\bibfnamefont {M.}~\bibnamefont
  {Sigrist}}\ and\ \bibinfo {author} {\bibfnamefont {D.~F.}\ \bibnamefont
  {Agterberg}},\ }\href@noop {} {\bibfield  {journal} {\bibinfo  {journal}
  {Prog. Theor. Phys.}\ }\textbf {\bibinfo {volume} {102}},\ \bibinfo {pages}
  {965} (\bibinfo {year} {1999})}\BibitemShut {NoStop}%
\bibitem [{\citenamefont {Babaev}\ \emph {et~al.}(2004)\citenamefont {Babaev},
  \citenamefont {Sudbo},\ and\ \citenamefont {Ashcroft}}]{Babaev04}%
  \BibitemOpen
  \bibfield  {author} {\bibinfo {author} {\bibfnamefont {E.}~\bibnamefont
  {Babaev}}, \bibinfo {author} {\bibfnamefont {A.}~\bibnamefont {Sudbo}}, \
  and\ \bibinfo {author} {\bibfnamefont {N.~W.}\ \bibnamefont {Ashcroft}},\
  }\href@noop {} {\bibfield  {journal} {\bibinfo  {journal} {Nature}\ }\textbf
  {\bibinfo {volume} {431}},\ \bibinfo {pages} {666} (\bibinfo {year}
  {2004})}\BibitemShut {NoStop}%
\bibitem [{\citenamefont {Volovik}\ and\ \citenamefont
  {Gor'kov}(1984)}]{Volovik84}%
  \BibitemOpen
  \bibfield  {author} {\bibinfo {author} {\bibfnamefont {G.~E.}\ \bibnamefont
  {Volovik}}\ and\ \bibinfo {author} {\bibfnamefont {L.~P.}\ \bibnamefont
  {Gor'kov}},\ }\href@noop {} {\bibfield  {journal} {\bibinfo  {journal} {JETP
  Lett.}\ }\textbf {\bibinfo {volume} {39}},\ \bibinfo {pages} {674} (\bibinfo
  {year} {1984})}\BibitemShut {NoStop}%
\bibitem [{\citenamefont {Volovik}(2000)}]{Volovik00}%
  \BibitemOpen
  \bibfield  {author} {\bibinfo {author} {\bibfnamefont {G.~E.}\ \bibnamefont
  {Volovik}},\ }\href@noop {} {\bibfield  {journal} {\bibinfo  {journal} {Proc.
  Nat. Acad. Sci. USA}\ }\textbf {\bibinfo {volume} {97}},\ \bibinfo {pages}
  {2431} (\bibinfo {year} {2000})}\BibitemShut {NoStop}%
\bibitem [{\citenamefont {Sigrist}\ \emph {et~al.}(1995)\citenamefont
  {Sigrist}, \citenamefont {Bailey},\ and\ \citenamefont
  {Laughlin}}]{Sigrist95}%
  \BibitemOpen
  \bibfield  {author} {\bibinfo {author} {\bibfnamefont {M.}~\bibnamefont
  {Sigrist}}, \bibinfo {author} {\bibfnamefont {D.~B.}\ \bibnamefont {Bailey}},
  \ and\ \bibinfo {author} {\bibfnamefont {R.~B.}\ \bibnamefont {Laughlin}},\
  }\href@noop {} {\bibfield  {journal} {\bibinfo  {journal} {Phys. Rev. Lett.}\
  }\textbf {\bibinfo {volume} {74}},\ \bibinfo {pages} {3249} (\bibinfo {year}
  {1995})}\BibitemShut {NoStop}%
\bibitem [{\citenamefont {Heeb}\ and\ \citenamefont
  {Agterberg}(1999)}]{Heeb99}%
  \BibitemOpen
  \bibfield  {author} {\bibinfo {author} {\bibfnamefont {R.}~\bibnamefont
  {Heeb}}\ and\ \bibinfo {author} {\bibfnamefont {D.~F.}\ \bibnamefont
  {Agterberg}},\ }\href@noop {} {\bibfield  {journal} {\bibinfo  {journal}
  {Phys. Rev. B}\ }\textbf {\bibinfo {volume} {59}},\ \bibinfo {pages} {7076}
  (\bibinfo {year} {1999})}\BibitemShut {NoStop}%
\bibitem [{\citenamefont {Mao}\ \emph {et~al.}(2000)\citenamefont {Mao},
  \citenamefont {Maeno}, \citenamefont {NishiZaki}, \citenamefont {Akima},\
  and\ \citenamefont {Ishiguro}}]{Mao00}%
  \BibitemOpen
  \bibfield  {author} {\bibinfo {author} {\bibfnamefont {Z.~Q.}\ \bibnamefont
  {Mao}}, \bibinfo {author} {\bibfnamefont {Y.}~\bibnamefont {Maeno}}, \bibinfo
  {author} {\bibfnamefont {S.}~\bibnamefont {NishiZaki}}, \bibinfo {author}
  {\bibfnamefont {T.}~\bibnamefont {Akima}}, \ and\ \bibinfo {author}
  {\bibfnamefont {T.}~\bibnamefont {Ishiguro}},\ }\href@noop {} {\bibfield
  {journal} {\bibinfo  {journal} {Phys. Rev. Lett.}\ }\textbf {\bibinfo
  {volume} {84}},\ \bibinfo {pages} {991} (\bibinfo {year} {2000})}\BibitemShut
  {NoStop}%
\bibitem [{\citenamefont {Deguchi}\ \emph {et~al.}(2004)\citenamefont
  {Deguchi}, \citenamefont {Mao}, \citenamefont {Yaguchi},\ and\ \citenamefont
  {Maeno}}]{Deguchi04}%
  \BibitemOpen
  \bibfield  {author} {\bibinfo {author} {\bibfnamefont {K.}~\bibnamefont
  {Deguchi}}, \bibinfo {author} {\bibfnamefont {Z.~Q.}\ \bibnamefont {Mao}},
  \bibinfo {author} {\bibfnamefont {H.}~\bibnamefont {Yaguchi}}, \ and\
  \bibinfo {author} {\bibfnamefont {Y.}~\bibnamefont {Maeno}},\ }\href@noop {}
  {\bibfield  {journal} {\bibinfo  {journal} {Phys. Rev. Lett.}\ }\textbf
  {\bibinfo {volume} {92}},\ \bibinfo {pages} {047002} (\bibinfo {year}
  {2004})}\BibitemShut {NoStop}%
\bibitem [{\citenamefont {Kittaka}\ \emph {et~al.}(2010)\citenamefont
  {Kittaka}, \citenamefont {Taniguchi}, \citenamefont {Yonezawa}, \citenamefont
  {Yaguchi},\ and\ \citenamefont {Maeno}}]{Kittaka10}%
  \BibitemOpen
  \bibfield  {author} {\bibinfo {author} {\bibfnamefont {S.}~\bibnamefont
  {Kittaka}}, \bibinfo {author} {\bibfnamefont {H.}~\bibnamefont {Taniguchi}},
  \bibinfo {author} {\bibfnamefont {S.}~\bibnamefont {Yonezawa}}, \bibinfo
  {author} {\bibfnamefont {H.}~\bibnamefont {Yaguchi}}, \ and\ \bibinfo
  {author} {\bibfnamefont {Y.}~\bibnamefont {Maeno}},\ }\href@noop {}
  {\bibfield  {journal} {\bibinfo  {journal} {Phys. Rev. B}\ }\textbf {\bibinfo
  {volume} {81}},\ \bibinfo {pages} {180510} (\bibinfo {year}
  {2010})}\BibitemShut {NoStop}%
\bibitem [{Note1()}]{Note1}%
  \BibitemOpen
  \bibinfo {note} {We do not need to specify whether the representation is
  $\Gamma _5^+$ or $\Gamma _5^-$, for which the basis functions transform
  respectively as $\protect \{XZ,YZ\protect \}$ or $\protect \{X,Y\protect \}$.
  The results of the present work apply to both cases.}\BibitemShut {Stop}%
\bibitem [{\citenamefont {Zhu}\ \emph {et~al.}(1997)\citenamefont {Zhu},
  \citenamefont {Ting}, \citenamefont {Shen},\ and\ \citenamefont
  {Wang}}]{Zhu97}%
  \BibitemOpen
  \bibfield  {author} {\bibinfo {author} {\bibfnamefont {J.-X.}\ \bibnamefont
  {Zhu}}, \bibinfo {author} {\bibfnamefont {C.~S.}\ \bibnamefont {Ting}},
  \bibinfo {author} {\bibfnamefont {J.~L.}\ \bibnamefont {Shen}}, \ and\
  \bibinfo {author} {\bibfnamefont {Z.~D.}\ \bibnamefont {Wang}},\ }\href@noop
  {} {\bibfield  {journal} {\bibinfo  {journal} {Phys. Rev. B}\ }\textbf
  {\bibinfo {volume} {56}},\ \bibinfo {pages} {14093} (\bibinfo {year}
  {1997})}\BibitemShut {NoStop}%
\bibitem [{\citenamefont {Bergemann}\ \emph {et~al.}(2000)\citenamefont
  {Bergemann}, \citenamefont {Julian}, \citenamefont {Mackenzie}, \citenamefont
  {NishiZaki},\ and\ \citenamefont {Maeno}}]{Bergemann00}%
  \BibitemOpen
  \bibfield  {author} {\bibinfo {author} {\bibfnamefont {C.}~\bibnamefont
  {Bergemann}}, \bibinfo {author} {\bibfnamefont {S.~R.}\ \bibnamefont
  {Julian}}, \bibinfo {author} {\bibfnamefont {A.~P.}\ \bibnamefont
  {Mackenzie}}, \bibinfo {author} {\bibfnamefont {S.}~\bibnamefont
  {NishiZaki}}, \ and\ \bibinfo {author} {\bibfnamefont {Y.}~\bibnamefont
  {Maeno}},\ }\href@noop {} {\bibfield  {journal} {\bibinfo  {journal} {Phys.
  Rev. Lett.}\ }\textbf {\bibinfo {volume} {84}},\ \bibinfo {pages} {2662}
  (\bibinfo {year} {2000})}\BibitemShut {NoStop}%
\bibitem [{\citenamefont {Haverkort}\ \emph {et~al.}(2008)\citenamefont
  {Haverkort}, \citenamefont {Elfimov}, \citenamefont {Tjeng}, \citenamefont
  {Sawatzky},\ and\ \citenamefont {Damascelli}}]{Haverkort08}%
  \BibitemOpen
  \bibfield  {author} {\bibinfo {author} {\bibfnamefont {M.~W.}\ \bibnamefont
  {Haverkort}}, \bibinfo {author} {\bibfnamefont {I.~S.}\ \bibnamefont
  {Elfimov}}, \bibinfo {author} {\bibfnamefont {L.~H.}\ \bibnamefont {Tjeng}},
  \bibinfo {author} {\bibfnamefont {G.~A.}\ \bibnamefont {Sawatzky}}, \ and\
  \bibinfo {author} {\bibfnamefont {A.}~\bibnamefont {Damascelli}},\
  }\href@noop {} {\bibfield  {journal} {\bibinfo  {journal} {Phys. Rev. Lett.}\
  }\textbf {\bibinfo {volume} {101}},\ \bibinfo {pages} {026406} (\bibinfo
  {year} {2008})}\BibitemShut {NoStop}%
\bibitem [{\citenamefont {Mermin}(1979)}]{Mermin79}%
  \BibitemOpen
  \bibfield  {author} {\bibinfo {author} {\bibfnamefont {N.~D.}\ \bibnamefont
  {Mermin}},\ }\href@noop {} {\bibfield  {journal} {\bibinfo  {journal} {Rev.
  Mod. Phys.}\ }\textbf {\bibinfo {volume} {51}},\ \bibinfo {pages} {591}
  (\bibinfo {year} {1979})}\BibitemShut {NoStop}%
\bibitem [{\citenamefont {Jackiw}\ and\ \citenamefont
  {Rebbi}(1976)}]{Jackiw76}%
  \BibitemOpen
  \bibfield  {author} {\bibinfo {author} {\bibfnamefont {R.}~\bibnamefont
  {Jackiw}}\ and\ \bibinfo {author} {\bibfnamefont {C.}~\bibnamefont {Rebbi}},\
  }\href@noop {} {\bibfield  {journal} {\bibinfo  {journal} {Phys. Rev. D}\
  }\textbf {\bibinfo {volume} {13}},\ \bibinfo {pages} {3398} (\bibinfo {year}
  {1976})}\BibitemShut {NoStop}%
\bibitem [{\citenamefont {Su}\ \emph {et~al.}(1980)\citenamefont {Su},
  \citenamefont {Schrieffer},\ and\ \citenamefont {Heeger}}]{Su80}%
  \BibitemOpen
  \bibfield  {author} {\bibinfo {author} {\bibfnamefont {W.~P.}\ \bibnamefont
  {Su}}, \bibinfo {author} {\bibfnamefont {J.~R.}\ \bibnamefont {Schrieffer}},
  \ and\ \bibinfo {author} {\bibfnamefont {A.~J.}\ \bibnamefont {Heeger}},\
  }\href@noop {} {\bibfield  {journal} {\bibinfo  {journal} {Phys. Rev. B}\
  }\textbf {\bibinfo {volume} {22}},\ \bibinfo {pages} {2099} (\bibinfo {year}
  {1980})}\BibitemShut {NoStop}%
\bibitem [{\citenamefont {Ran}\ \emph {et~al.}(2009)\citenamefont {Ran},
  \citenamefont {Zhang},\ and\ \citenamefont {Vishwanath}}]{Ran09}%
  \BibitemOpen
  \bibfield  {author} {\bibinfo {author} {\bibfnamefont {Y.}~\bibnamefont
  {Ran}}, \bibinfo {author} {\bibfnamefont {Y.}~\bibnamefont {Zhang}}, \ and\
  \bibinfo {author} {\bibfnamefont {A.}~\bibnamefont {Vishwanath}},\
  }\href@noop {} {\bibfield  {journal} {\bibinfo  {journal} {Nat. Phys.}\
  }\textbf {\bibinfo {volume} {5}},\ \bibinfo {pages} {298} (\bibinfo {year}
  {2009})}\BibitemShut {NoStop}%
\bibitem [{Note2()}]{Note2}%
  \BibitemOpen
  \bibinfo {note} {This choice of parametrization is similar to that used in
  Ref.~\cite {Sigrist99}, in which the aforementioned additional structure of
  the order is parametrized by the scalar fields $\alpha $ and $\chi $ via
  ${\unhbox \voidb@x \hbox {\protect \boldmath $\eta $}} \propto {\unhbox
  \voidb@x \hbox {\protect \boldmath $R$}}^{-\alpha / 2} \cdot {\setbox \z@
  \hbox {\frozen@everymath \@emptytoks \mathsurround \z@ $\nulldelimiterspace
  \z@ \left (\vcenter to\@ne \big@size {}\right .$}\box \z@ } \protect \qopname
  \relax o{cos}(\chi +\pi /4),i \protect \qopname \relax o{sin}(\chi +\pi
  /4){\setbox \z@ \hbox {\frozen@everymath \@emptytoks \mathsurround \z@
  $\nulldelimiterspace \z@ \left )\vcenter to\@ne \big@size {}\right .$}\box
  \z@ }$.}\BibitemShut {Stop}%
\bibitem [{Note3()}]{Note3}%
  \BibitemOpen
  \bibinfo {note} {We note that for weak coupling and specular reflection, the
  condition at the sample boundary requires $\beta =0$ (see, e.g., Ref.~\cite
  {Sigrist91}), and thus, for a finite sample, all regions can be considered to
  be surrounded by domain-wall loops. However, the coupling between $\Gamma $
  and the normal direction of either the surface or the domain wall is
  generically different, which can result in distinct equilibrium orientations
  of $\Gamma $ for each of these cases (see, e.g. ~\cite
  {Matsumoto99}).}\BibitemShut {Stop}%
\bibitem [{\citenamefont {Izyumov}\ and\ \citenamefont
  {Laptev}(1990)}]{Izyumov90}%
  \BibitemOpen
  \bibfield  {author} {\bibinfo {author} {\bibfnamefont {Y.~A.}\ \bibnamefont
  {Izyumov}}\ and\ \bibinfo {author} {\bibfnamefont {V.~M.}\ \bibnamefont
  {Laptev}},\ }\href@noop {} {\bibfield  {journal} {\bibinfo  {journal} {Phase
  Transitions}\ }\textbf {\bibinfo {volume} {20}},\ \bibinfo {pages} {95}
  (\bibinfo {year} {1990})}\BibitemShut {NoStop}%
\bibitem [{\citenamefont {Tokuyasu}\ \emph {et~al.}(1990)\citenamefont
  {Tokuyasu}, \citenamefont {Hess},\ and\ \citenamefont {Sauls}}]{Tokuyasu90}%
  \BibitemOpen
  \bibfield  {author} {\bibinfo {author} {\bibfnamefont {T.~A.}\ \bibnamefont
  {Tokuyasu}}, \bibinfo {author} {\bibfnamefont {D.~W.}\ \bibnamefont {Hess}},
  \ and\ \bibinfo {author} {\bibfnamefont {J.~A.}\ \bibnamefont {Sauls}},\
  }\href@noop {} {\bibfield  {journal} {\bibinfo  {journal} {Phys. Rev. B}\
  }\textbf {\bibinfo {volume} {41}},\ \bibinfo {pages} {8891} (\bibinfo {year}
  {1990})}\BibitemShut {NoStop}%
\bibitem [{\citenamefont {Sauls}\ and\ \citenamefont {Eschrig}()}]{Sauls09}%
  \BibitemOpen
  \bibfield  {author} {\bibinfo {author} {\bibfnamefont {J.~A.}\ \bibnamefont
  {Sauls}}\ and\ \bibinfo {author} {\bibfnamefont {M.}~\bibnamefont
  {Eschrig}},\ }\href@noop {} {\bibfield  {journal} {\bibinfo  {journal} {N.
  Jo. Phys.}\ }\textbf {\bibinfo {volume} {11}}}\BibitemShut {NoStop}%
\bibitem [{\citenamefont {Ichioka}\ \emph {et~al.}(2005)\citenamefont
  {Ichioka}, \citenamefont {Matsunaga},\ and\ \citenamefont
  {Machida}}]{Ichioka05}%
  \BibitemOpen
  \bibfield  {author} {\bibinfo {author} {\bibfnamefont {M.}~\bibnamefont
  {Ichioka}}, \bibinfo {author} {\bibfnamefont {Y.}~\bibnamefont {Matsunaga}},
  \ and\ \bibinfo {author} {\bibfnamefont {K.}~\bibnamefont {Machida}},\
  }\href@noop {} {\bibfield  {journal} {\bibinfo  {journal} {Phys. Rev. B}\
  }\textbf {\bibinfo {volume} {71}},\ \bibinfo {pages} {172510} (\bibinfo
  {year} {2005})}\BibitemShut {NoStop}%
\bibitem [{\citenamefont {Polyakov}(1987)}]{Polyakov87}%
  \BibitemOpen
  \bibfield  {author} {\bibinfo {author} {\bibfnamefont {A.~M.}\ \bibnamefont
  {Polyakov}},\ }\href@noop {} {\emph {\bibinfo {title} {Gauge Fields and
  Strings}}}\ (\bibinfo  {publisher} {Harwood Academic Publishers},\ \bibinfo
  {address} {New York},\ \bibinfo {year} {1987})\BibitemShut {NoStop}%
\bibitem [{\citenamefont {Zee}(2003)}]{Zee03}%
  \BibitemOpen
  \bibfield  {author} {\bibinfo {author} {\bibfnamefont {A.}~\bibnamefont
  {Zee}},\ }\href@noop {} {\emph {\bibinfo {title} {Quantum Field Theory in a
  Nutshell}}}\ (\bibinfo  {publisher} {Princeton University Press},\ \bibinfo
  {address} {Princeton},\ \bibinfo {year} {2003})\BibitemShut {NoStop}%
\bibitem [{Note4()}]{Note4}%
  \BibitemOpen
  \bibinfo {note} {Here and elsewhere in this paper, we take into account the
  core energies of domain walls but not the core energies of vortices. Our
  justification for doing this is that, in the standard London limit, the
  energy cost of a vortex core is negligibly small, compared with the kinetic
  energy of the supercurrents and magnetic fields, whereas the core energy of a
  domain wall is not.}\BibitemShut {Stop}%
\bibitem [{Note5()}]{Note5}%
  \BibitemOpen
  \bibinfo {note} {In the case that the equilibrium value of $\Gamma - \phi $
  is $\pi /2$, the dipole part of the vorticity is given by ${\unhbox \voidb@x
  \hbox {\protect \boldmath $d$}}={\setbox \z@ \hbox {\frozen@everymath
  \@emptytoks \mathsurround \z@ $\nulldelimiterspace \z@ \left (\vcenter to\@ne
  \big@size {}\right .$}\box \z@ }-(\pi \tau /4)-\mu {\setbox \z@ \hbox
  {\frozen@everymath \@emptytoks \mathsurround \z@ $\nulldelimiterspace \z@
  \left )\vcenter to\@ne \big@size {}\right .$}\box \z@ }{\unhbox \voidb@x
  \hbox {\protect \boldmath $I$}}$. Thus, as is shown in the referring
  paragraph, the part of the domain-wall current proportional to $\tau $ can
  flow in either direction, for a given pattern of chirality, depending on the
  value of $\Gamma - \phi $.}\BibitemShut {Stop}%
\bibitem [{\citenamefont {Kwon}\ \emph {et~al.}(2003)\citenamefont {Kwon},
  \citenamefont {Yakovenko},\ and\ \citenamefont {Sengupta}}]{Kwon03}%
  \BibitemOpen
  \bibfield  {author} {\bibinfo {author} {\bibfnamefont {H.-J.}\ \bibnamefont
  {Kwon}}, \bibinfo {author} {\bibfnamefont {V.~M.}\ \bibnamefont {Yakovenko}},
  \ and\ \bibinfo {author} {\bibfnamefont {K.}~\bibnamefont {Sengupta}},\
  }\href@noop {} {\bibfield  {journal} {\bibinfo  {journal} {Synthetic Metals}\
  }\textbf {\bibinfo {volume} {133-134}},\ \bibinfo {pages} {27} (\bibinfo
  {year} {2003})}\BibitemShut {NoStop}%
\bibitem [{Note6()}]{Note6}%
  \BibitemOpen
  \bibinfo {note} {For example, as noted in Ref.~\cite {Volovik85}, in the
  limit of large domain-wall currents, vortices may be stabilized along the
  domain wall. Such vortices would then spoil the translational invariance of
  the domain wall. Although we know of no experimental evidence for such an
  effect, there is evidence from Josephson-junction tunneling experiments on
  Sr$_2$RuO$_4$~\cite {Kidwingira06} for the absence of translational
  invariance (i.e., for the presence of multiple domains) along sample
  boundaries, which can be regarded in certain respects as analogous to domain
  walls.}\BibitemShut {Stop}%
\bibitem [{\citenamefont {Geshkenbein}\ \emph {et~al.}(1987)\citenamefont
  {Geshkenbein}, \citenamefont {Larkin},\ and\ \citenamefont
  {Barone}}]{Geshkenbein87}%
  \BibitemOpen
  \bibfield  {author} {\bibinfo {author} {\bibfnamefont {V.~B.}\ \bibnamefont
  {Geshkenbein}}, \bibinfo {author} {\bibfnamefont {A.~I.}\ \bibnamefont
  {Larkin}}, \ and\ \bibinfo {author} {\bibfnamefont {A.}~\bibnamefont
  {Barone}},\ }\href@noop {} {\bibfield  {journal} {\bibinfo  {journal} {Phys.
  Rev. B}\ }\textbf {\bibinfo {volume} {36}},\ \bibinfo {pages} {235} (\bibinfo
  {year} {1987})}\BibitemShut {NoStop}%
\bibitem [{Note7()}]{Note7}%
  \BibitemOpen
  \bibinfo {note} {As a particular case of Eq.~(\ref {eqn:Bend_Angle_Flux}),
  one can consider a straight domain wall. Then, Eqs.~(\ref {eqn:change-alpha})
  and (\ref {eqn:Bend_Angle_Flux}) imply that the topologically stable,
  localized solitons in $\Gamma (s)$ along a domain wall would obey $\Delta
  \Gamma = \pi $, and that each is associated with a flux $\Phi _0$. However,
  if the dependence of $E_{\protect \rm dw}$ on $\Gamma (s)-\phi (s)$ should
  have multiple minima per $\pi $ period, then there could be topologically
  stable solitons in $\Gamma (s)$ along a straight domain wall, each having
  $\Delta \Gamma \not =\pi $ and connected with nonquantized amounts of flux
  (see Refs.~\cite {Sigrist89,Sigrist99,Bouhon10}). In this case, in addition
  to the bend flux of Eq.~(\ref {eqn:Bend_Angle_Flux}), the flux associated
  with a bent domain wall may have a further contribution.}\BibitemShut {Stop}%
\bibitem [{\citenamefont {Landau}\ and\ \citenamefont
  {Lifshitz}(1977)}]{Landau77}%
  \BibitemOpen
  \bibfield  {author} {\bibinfo {author} {\bibfnamefont {L.~D.}\ \bibnamefont
  {Landau}}\ and\ \bibinfo {author} {\bibfnamefont {E.~M.}\ \bibnamefont
  {Lifshitz}},\ }\href@noop {} {\emph {\bibinfo {title} {Quantum Mechanics :
  Non-relativistic Theory}}},\ \bibinfo {edition} {3rd}\ ed.\ (\bibinfo
  {publisher} {Pergamon Press},\ \bibinfo {address} {New York},\ \bibinfo
  {year} {1977})\BibitemShut {NoStop}%
\bibitem [{\citenamefont {Leggett}(2006)}]{Leggett06}%
  \BibitemOpen
  \bibfield  {author} {\bibinfo {author} {\bibfnamefont {A.~J.}\ \bibnamefont
  {Leggett}},\ }\href@noop {} {\emph {\bibinfo {title} {Quantum Liquids : Bose
  Condensation and Cooper Pairing in Condensed-Matter Systems}}}\ (\bibinfo
  {publisher} {Oxford University Press},\ \bibinfo {address} {New York},\
  \bibinfo {year} {2006})\BibitemShut {NoStop}%
\bibitem [{\citenamefont {Ashby}\ and\ \citenamefont {Kallin}(2009)}]{Ashby09}%
  \BibitemOpen
  \bibfield  {author} {\bibinfo {author} {\bibfnamefont {P.~E.~C.}\
  \bibnamefont {Ashby}}\ and\ \bibinfo {author} {\bibfnamefont
  {C.}~\bibnamefont {Kallin}},\ }\href@noop {} {\bibfield  {journal} {\bibinfo
  {journal} {Phys. Rev. B}\ }\textbf {\bibinfo {volume} {79}},\ \bibinfo
  {pages} {224509} (\bibinfo {year} {2009})}\BibitemShut {NoStop}%
\bibitem [{\citenamefont {Bluhm}(2007)}]{Bluhm07}%
  \BibitemOpen
  \bibfield  {author} {\bibinfo {author} {\bibfnamefont {H.}~\bibnamefont
  {Bluhm}},\ }\href@noop {} {\bibfield  {journal} {\bibinfo  {journal} {Phys.
  Rev. B}\ }\textbf {\bibinfo {volume} {76}},\ \bibinfo {pages} {144507}
  (\bibinfo {year} {2007})}\BibitemShut {NoStop}%
\bibitem [{\citenamefont {Dolocan}\ \emph {et~al.}(2005)\citenamefont
  {Dolocan}, \citenamefont {Veauvy}, \citenamefont {Servant}, \citenamefont
  {Lejay}, \citenamefont {Hasselbach}, \citenamefont {Liu},\ and\ \citenamefont
  {Mailly}}]{Dolocan05}%
  \BibitemOpen
  \bibfield  {author} {\bibinfo {author} {\bibfnamefont {V.~O.}\ \bibnamefont
  {Dolocan}}, \bibinfo {author} {\bibfnamefont {C.}~\bibnamefont {Veauvy}},
  \bibinfo {author} {\bibfnamefont {F.}~\bibnamefont {Servant}}, \bibinfo
  {author} {\bibfnamefont {P.}~\bibnamefont {Lejay}}, \bibinfo {author}
  {\bibfnamefont {K.}~\bibnamefont {Hasselbach}}, \bibinfo {author}
  {\bibfnamefont {Y.}~\bibnamefont {Liu}}, \ and\ \bibinfo {author}
  {\bibfnamefont {D.}~\bibnamefont {Mailly}},\ }\href@noop {} {\bibfield
  {journal} {\bibinfo  {journal} {Phys. Rev. Lett.}\ }\textbf {\bibinfo
  {volume} {95}},\ \bibinfo {pages} {097004} (\bibinfo {year}
  {2005})}\BibitemShut {NoStop}%
\bibitem [{\citenamefont {Dolocan}\ \emph {et~al.}(2006)\citenamefont
  {Dolocan}, \citenamefont {Lejay}, \citenamefont {Mailly},\ and\ \citenamefont
  {Hasselbach}}]{Dolocan06}%
  \BibitemOpen
  \bibfield  {author} {\bibinfo {author} {\bibfnamefont {V.~O.}\ \bibnamefont
  {Dolocan}}, \bibinfo {author} {\bibfnamefont {P.}~\bibnamefont {Lejay}},
  \bibinfo {author} {\bibfnamefont {D.}~\bibnamefont {Mailly}}, \ and\ \bibinfo
  {author} {\bibfnamefont {K.}~\bibnamefont {Hasselbach}},\ }\href@noop {}
  {\bibfield  {journal} {\bibinfo  {journal} {Phys. Rev. B}\ }\textbf {\bibinfo
  {volume} {74}},\ \bibinfo {pages} {144505} (\bibinfo {year}
  {2006})}\BibitemShut {NoStop}%
\bibitem [{\citenamefont {Logoboy}\ and\ \citenamefont
  {Sonin}(2009)}]{Logoboy09}%
  \BibitemOpen
  \bibfield  {author} {\bibinfo {author} {\bibfnamefont {N.~A.}\ \bibnamefont
  {Logoboy}}\ and\ \bibinfo {author} {\bibfnamefont {E.~B.}\ \bibnamefont
  {Sonin}},\ }\href@noop {} {\bibfield  {journal} {\bibinfo  {journal} {Phys.
  Rev. B}\ }\textbf {\bibinfo {volume} {79}},\ \bibinfo {pages} {094511}
  (\bibinfo {year} {2009})}\BibitemShut {NoStop}%
\bibitem [{\citenamefont {Bouhon}\ and\ \citenamefont
  {Sigrist}(2010)}]{Bouhon10}%
  \BibitemOpen
  \bibfield  {author} {\bibinfo {author} {\bibfnamefont {A.}~\bibnamefont
  {Bouhon}}\ and\ \bibinfo {author} {\bibfnamefont {M.}~\bibnamefont
  {Sigrist}},\ }\href@noop {} {\bibfield  {journal} {\bibinfo  {journal} {N.
  Jo. Phys.}\ }\textbf {\bibinfo {volume} {12}},\ \bibinfo {pages} {043031}
  (\bibinfo {year} {2010})}\BibitemShut {NoStop}%
\bibitem [{\citenamefont {Kidwingira}\ \emph {et~al.}(2006)\citenamefont
  {Kidwingira}, \citenamefont {Strand}, \citenamefont {Harlingen},\ and\
  \citenamefont {Maeno}}]{Kidwingira06}%
  \BibitemOpen
  \bibfield  {author} {\bibinfo {author} {\bibfnamefont {F.}~\bibnamefont
  {Kidwingira}}, \bibinfo {author} {\bibfnamefont {J.~D.}\ \bibnamefont
  {Strand}}, \bibinfo {author} {\bibfnamefont {D.~J.~V.}\ \bibnamefont
  {Harlingen}}, \ and\ \bibinfo {author} {\bibfnamefont {Y.}~\bibnamefont
  {Maeno}},\ }\href@noop {} {\bibfield  {journal} {\bibinfo  {journal}
  {Science}\ }\textbf {\bibinfo {volume} {314}},\ \bibinfo {pages} {1267}
  (\bibinfo {year} {2006})}\BibitemShut {NoStop}%
\end{thebibliography}%

\end{document}